%% file: bbflow.tex
\documentclass[aps,showpacs,nofootinbib,prd,twocolumn,superscriptaddress,floatfix]{revtex4-1}
\usepackage[utf8]{inputenc}
\usepackage{graphicx}
\usepackage{amssymb}
\usepackage{amsmath}
\usepackage{xcolor}
\usepackage{mathtools,slashed}
\usepackage{epstopdf}
\usepackage{float}
\usepackage{hyperref}
\usepackage{MnSymbol}
\hypersetup{colorlinks,citecolor=blue,urlcolor=blue,linkcolor=black}

\def\MSBAR{{\overline{\mathrm{MS}}}}
\def\mubar{\bar{\mu}}
\def\muUV{\bar{\mu}_{\mathrm{UV}}}
\def\muIR{\bar{\mu}_{\mathrm{IR}}}
\def\tauf{{\tau_{\mathrm{F}}}}

\def\Tr{\,\mathrm{Tr}\:}
\def\Eq#1{Eq.~(\ref{#1})}
\def\GE{G_{\mathrm{E}}}
\def\GB{G_{\mathrm{B}}}
\def\GBphysical{G_{\mathrm{B}}^{\mathrm{phys}}}
\def\CR{\mathcal{C}_{\mathcal{R}}}
\def\DR{d_{\mathcal{R}}}
\def\TR{T_{\mathcal{R}}}
\def\CA{\mathcal{C}_{\mathcal{A}}}
\def\kpp{k_0^{\mathrm{pp}}}

\begin{document}


\title{
QCD field-strength correlators on a Polyakov loop with gradient flow at next-to-leading order
}

\author{David de la Cruz}
\affiliation{Institut f\"ur Kernphysik, Technische Universit\"at Darmstadt\\
Schlossgartenstra{\ss}e 2, D-64289 Darmstadt, Germany}

\author{Alexander M. Eller}
\affiliation{Institut f\"ur Kernphysik, Technische Universit\"at Darmstadt\\
Schlossgartenstra{\ss}e 2, D-64289 Darmstadt, Germany}

\author{Guy D. Moore}
\affiliation{Institut f\"ur Kernphysik, Technische Universit\"at Darmstadt\\
Schlossgartenstra{\ss}e 2, D-64289 Darmstadt, Germany}

\begin{abstract}
Momentum exchange between a heavy quark and a hot quark-gluon medium can be characterized nonperturbatively in terms of field-strength field-strength ($E-E$ and $B-B$) correlators along a Polyakov loop.
These can be studied on the lattice and analytically continued.
However the lattice typically determines the correlators after the application of gradient flow.
We investigate how gradient flow renormalizes these correlation functions by carrying out a next-to-leading order perturbative analysis of the correlators including gradient flow.
This establishes a next-to-leading order renormalization matching between the correlators as measured on the lattice and the correlators relevant for momentum diffusion.


\end{abstract}

\maketitle

\section{Introduction}
\label{sec:introduction}

Relativistic collisions of large ions create a nearly-equilibrium Quark-Gluon Plasma \cite{koch_event-by-event_2002,STAR:2005gfr,arsene_quarkgluon_2005,adcox_formation_2005,na49_collaboration_pion_2008,PhysRevLett.103.251601,jet_collaboration_extracting_2014}
which we would like to characterize in as many ways as possible.
Heavy quarks represent one of the most interesting observables or probes of such collisions.
The large mass of the quark simplifies the description of heavy quark interactions with the medium of light species and gives the heavy quarks a nearly-conserved quantum number which makes them easier to track through the plasma.
The behavior of heavy quarks varies depending on their energy range.
The highest-energy heavy quarks form jets, and the medium modification of these jets differs from that for light-quark or gluon jets because the quark's mass introduces the so-called dead-cone effect
\cite{dokshitzer_specific_1991,thomas_gluon_2005,maltoni_exposing_2016,acharya_direct_2022,kluth_observation_2023}.
Experimental observations have confirmed this distinction from light quark jets \cite{khoze_perturbative-qcd_1997,abreu_hadronization_2000,dokshitzer_heavy-quark_2001}.
The intermediate energy regime is more complex, as both elastic and radiative energy loss mechanisms are expected to play a role
\cite{phenix_collaboration_suppression_2001,xiang_charm_2005,armesto_testing_2005,bielcik_centrality_2006,renk_phenomenology_2007,wicks_elastic_2007,phenix_collaboration_energy_2007,majumder_elastic_2009,uphoff_open_2012,berrehrah_dynamical_2014}.

This paper will focus on the behavior of relatively low energy heavy quarks in a medium.
Because the quark mass is heavy, the typical momentum is large compared to the thermal scale, $|\vec p|^2 \sim 3MT \gg T^2$.
It therefore takes many independent particle-medium interactions for the momentum to change significantly.
This gives rise to Langevin dynamics for the momentum
\cite{Svetitsky:1987gq}, characterized by a momentum diffusion coefficient $\kappa$.
This problem has already been addressed rather extensively.
We know that the convergence of perturbation theory for heavy-quark diffusion is particularly poor
\cite{Caron-Huot:2007rwy}, calling for nonperturbative lattice approaches.
A nonperturbative definition of momentum diffusion was established in Ref.~\cite{casalderrey-solana_heavy_2006} in terms of an electric field correlation function along a Wilson line running along the time direction.
Ref.~\cite{Caron-Huot:2009ncn} showed how to relate this correlator to a Euclidean-space correlator along a Polyakov loop, allowing its extraction via analytical continuation of lattice data.
More recently, Ref.~\cite{Bouttefeux:2020ycy} showed that finite-mass corrections are captured by a correlator of magnetic fields on a Polyakov loop.
A number of lattice studies of these correlators have already taken place in an attempt to exploit this possibility of gaining nonperturbative dynamical information about heavy quark diffusion
\cite{Meyer:2010tt,Francis:2011gc,Banerjee:2011ra,Kaczmarek:2014jga,Neuhaus:2015qja,Francis:2015daa,Brambilla:2019oaa,Altenkort:2020fgs,Brambilla:2020siz,Mayer-Steudte:2021hei,Banerjee:2022uge,Brambilla:2022xbd,Banerjee:2022gen,Altenkort:2023oms,Altenkort:2023eav,Altenkort:2024spl}.

The principal challenge of such lattice studies is to achieve a good signal-to-noise ratio by suppressing short-distance fluctuations.
The modern approach to doing this is through the use of gradient flow
\cite{luscher_properties_2010,Christensen:2016wdo,Altenkort:2020fgs,Brambilla:2022xbd}.
This approach provides automatically UV-renormalized operators, suppresses noise, and is straightforward to use in full (unquenched) QCD.
The disadvantage is the need to understand the relation between the renormalized operators provided by gradient flow and the operators which define the heavy quark diffusion.

The goal of this paper is precisely to understand the issues of UV renormalization for electric and magnetic field strengths inserted on Wilson lines under gradient flow at next-to-leading order (NLO) in the strong coupling, for applications to heavy quark diffusion.
This technical achievement can then be applied in lattice studies of heavy quark diffusion to provide correct and accurate matching between the measured operators and the correlators which are physically relevant for heavy quark diffusion.
In the next section we will review heavy quark diffusion and electric and magnetic correlators on Polyakov loops.
Next, Section \ref{sec:gradientflow} reviews the gradient flow formalism and its Feynman rules.
Then Section \ref{sec:diagrams} presents the relevant Feynman diagrams and their evaluation.
Section \ref{sec:results} presents the diagram-by-diagram results for both electric and magnetic flow correlation functions along a vacuum, fundamental Wilson line.
Section \ref{sec:combining} combines the results to establish both the NLO vacuum spectral function and the NLO matching between the Wilson-line gauge field correlator under flow and the same correlator without flow in the $\MSBAR$ renormalization scheme.
A few appendices present some further details.
Feynman rules are reviewed in Appendix \ref{secFeynRules}, individual expressions for diagrams are given in Appendix \ref{secDetails}, and the results if the fundamental Wilson line is replaced by an adjoint Wilson line running only between the two field insertions are discussed in Appendix \ref{secadjointline}.

Before the completion of this work, its most important results were already presented and applied to both a quenched
\cite{Altenkort:2024spl}
and an unquenched
\cite{Altenkort:2023eav}
study of mass-dependent heavy quark diffusion.
Also, recently another study of operator renormalization on a Wilson line with gradient flow appeared, which agrees with our results for the renormalization factor for both electric and magnetic fields
\cite{Brambilla:2023vwm},
though that study does not provide a calculation of the NLO spectral function.
A previous calculation of the NLO spectral function without flow in the $\MSBAR$ scheme, both in vacuum and including finite-temperature effects, has been provided in Ref.~\cite{Burnier:2010rp} for the electric field case and in Ref.~\cite{Laine:2021uzs} for the magnetic field case.  
Our results agree with these references except for the contribution of diagram $(j)$, leading to a different coefficient on the $\pi^2$ term in the NLO result.
The same difference was also recently found in Ref.~\cite{Scheihing_Hitschfeld_2023}, whose Appendix C explains the computational error of Ref.~\cite{Burnier:2010rp}.

\section{Review of heavy quark diffusion}
\label{sec:diffusion}

This section provides an intuitive overview of heavy quark diffusion and its relationship to electric and magnetic field correlators along Wilson lines.
Nothing in this section is original, but it helps to set the context and pose the problem for the rest of the paper.

We assume a separation between the scale $\sim 3MT$ of momentum-squared a typical heavy quark carries and the scale $\sim T^2$ of momentum-squared which is transferred to a heavy quark in a coherent medium interaction.
This is the same as an $M \gg T$ expansion.
Within this approximation, the momentum diffusion coefficient can be defined as the mean squared momentum transferred per unit time to each vector momentum component,
\begin{equation}
\label{defkappa}
    \frac 13 \lim_{\Delta t \gg 1/T} \frac{\langle |\Delta \vec p|^2 \rangle}{\Delta t}
    \equiv \kappa .
\end{equation}
Here $\Delta t$ is understood as a time scale long compared to the coherence time of medium interactions, but small compared to the time it takes to transfer $(\Delta p)^2 \sim 3MT$.
In the approximation that a heavy quark's momentum-space evolution is controlled by a Langevin equation, this coefficient determines the magnitude of the ``noise'' term:
\begin{equation}
\label{Langevin}
    \frac{d\, p_i(t)}{d t} = - \eta_D p_i(t) + \xi_i(t) \,,
    \quad
    \langle \xi_i(t) \xi_j(t') \rangle 
    = \kappa \delta_{ij} \delta(t-t') \,.
\end{equation}
The momentum-diffusion and drag coefficients are related by an Einstein relation:
\begin{equation}
\label{Einstein}
    \eta_D = \frac{\kappa}{2MT}.
\end{equation}
The spatial diffusion coefficient $D_s$ is also related to $\kappa$ via
$D_s = 2 T^2/\kappa$.
Since these quantities are all related, it is conventional to focus on $\kappa$, and we will do so here as well.

The momentum change of a heavy quark is the force acting on the heavy quark, $d\vec p/dt = \vec F$, which allows us to relate $\kappa$ to a force-force correlator:
\begin{align}
\label{forceforce}
    \Delta \vec p & = \int_0^{t} \vec F(t') \nonumber \\
    \frac{1}{3t} (\Delta \vec p)^2 & = \frac{1}{3t}
    \int_0^t dt' \int_0^t dt'' \vec F(t') \cdot \vec F(t'') \nonumber \\
    \lim_{t \gg 1/T} \frac{1}{3t} \langle (\Delta p)^2 \rangle & \simeq \frac 13
    \int_{-\infty}^{\infty} dt' \langle \vec F(t') \cdot \vec F(0) \rangle
\end{align}
where in the last step we used that the $\vec F(t') \cdot \vec F(t'')$ correlator is time-translation invariant and we extended the integration range because the correlator is dominated by times $t' \sim 1/T$.
This relation forms the basis for perturbative analyses of $\kappa$.
Perturbative calculations of $\kappa$
\cite{Svetitsky:1987gq,Moore:2004tg} have been extended to next-to-leading order in weak coupling
\cite{Caron-Huot:2007rwy},
with the uncomfortable conclusion that the perturbative expansion works very poorly.
However, Casalderrey-Solana and Teaney have shown how to define the heavy quark diffusion coefficient in a rigorous and nonperturbative way
\cite{casalderrey-solana_heavy_2006} in terms of the correlation function of electric fields along a Wilson line, representing the propagation of the heavy quark and making the calculation gauge invariant.
Defining the electric field correlator
\begin{equation}
\label{Wightman}
    G^>(t) = \frac{1}{3} \frac{\langle \Tr U(0,t) E_j(t) U(t,0) E_j(0)U(0,-i\beta) \rangle}{\langle \Tr U(0,-i\beta) \rangle }
\end{equation}
with $\langle .. \rangle$ representing averaging in a thermal ensemble,
$U(t_1,t_2)$ a fundamental-representation Wilson line at a fixed space point along the time contour from $t_1$ to $t_2$, $E_j=F_{0j}$ the electric field in direction $j$, and $U(0,-i\beta)$ the Wilson line across the thermal Euclidean extent (Polyakov loop), we have
\begin{equation}
\label{kappaWightman}
    \kappa_E = \int_{-\infty}^{\infty} G^>(t) dt \,.
\end{equation}
Alternatively we can consider the corresponding spectral function $\rho(\omega) = i \int e^{i\omega t} (G^>(t)-G^>(-t)) dt$ and write $\kappa = \lim_{\omega \to 0} (2T/\omega) \rho(\omega)$.

Caron-Huot, Laine, and Moore have shown
\cite{Caron-Huot:2009ncn}
that the correlator in \Eq{Wightman} is the analytical continuation of the Euclidean-time correlator
\begin{equation}
    \label{Euclidean}
    \GE(\tau) = -\frac{1}{3} \frac{\langle \Tr E_j(0) U(0,-i\tau) E_j(-i\tau) U(-i\tau,-i\beta) \rangle}{\langle\Tr U(0,-i\beta)\rangle}.
\end{equation}
Note that $E_i = F_{0i}$ is now the Euclidean field strength -- the time direction is the Euclidean time direction -- which differs from the Minkowski one by an imaginary factor, leading to the overall minus sign.
Caron-Huot, Laine, and Moore also showed that the $\tau$ dependence of this correlator has the same relation to the spectral function $\rho(\omega)$ as for the two-point function of a gauge-invariant operator:
\begin{equation}
    \label{continuation}
    \GE(\tau) = \int_{-\infty}^{\infty} \frac{d\omega}{2\pi}
    \rho_E(\omega) \frac{\cosh \omega(\tau-\beta/2)}{\sinh \beta\omega/2}.
\end{equation}
This opened the possibility of computing $\kappa$ using nonperturbative lattice simulations to determine $\GE(\tau)$ and analytical continuation to reconstruct $\rho(\omega)$ and particularly its $\omega \to 0$ limit: $\lim_{\omega \to 0} \rho_E(\omega) / \omega = \kappa_E$.
Many lattice studies based on this approach have followed
\cite{Meyer:2010tt,Francis:2011gc,Banerjee:2011ra,Kaczmarek:2014jga,Neuhaus:2015qja,Francis:2015daa,Brambilla:2019oaa,Altenkort:2020fgs,Brambilla:2020siz,Mayer-Steudte:2021hei,Banerjee:2022uge,Brambilla:2022xbd,Banerjee:2022gen,Altenkort:2023oms,Altenkort:2023eav,Altenkort:2024spl}.

Recently Bouttefeux and Laine have shown that finite-velocity or subleading-mass corrections can also be investigated
\cite{Bouttefeux:2020ycy} by studying the correlation function of two \textsl{magnetic} field operators on a Polyakov loop,
\begin{equation}
    \label{BBEuclidean}
    \GB(\tau) = \frac{1}{6} \frac{\langle \Tr F_{jk}(0) U(0,i\tau) F_{jk}(i\tau) U(i\tau,i\beta) \rangle}{\langle\Tr U(0,i\beta) \rangle}
\end{equation}
where $F_{jk}$ is the magnetic field in the $(jk)$ plane, and the factor of $1/6$ would be $1/3$ if we restricted to $j>k$ to capture each magnetic field component only once.
The associated coefficient $\kappa_B$, defined in analogy to $\kappa_E$, determines a subleading-in-velocity contribution to the momentum diffusion:
\begin{equation}
\label{fullkappa}
    \kappa = \kappa_E + \frac 23 \langle v^2 \rangle \kappa_B \,,
\end{equation}
where $\langle v^2 \rangle$ is the mass-dependent mean-squared velocity of the heavy quark.
The factor of $v^2$ arises because the force from a magnetic field is proportional to velocity and the 2/3 factor arises because only the two transverse magnetic field components exert a force on a moving charge.
Note that applying \Eq{BBEuclidean} at NLO in the coupling requires a careful matching between the renormalized operator and the physical force on a charged particle.
Fortunately, this matching was carried out at NLO in Ref.~\cite{Laine:2021uzs}, matching however to the $\MSBAR$ scheme without gradient flow.

\section{Gradient flow and its Feynman rules}
\label{sec:gradientflow}

The relation between forces and electromagnetic fields on the Wilson line is intuitively clear, but it does not automatically extend beyond leading order in the coupling.
Indeed, both the Wilson line, the field strength, and the way they attach to each other are subject to renormalization.
The (perimeter-law, UV divergent) renormalization of the Wilson line itself is controlled by the denominator factor in \Eq{continuation} and \Eq{BBEuclidean},
but renormalization of the operator insertion on the Wilson line is not.
As already indicated, modern lattice techniques compute the correlators of \Eq{continuation} and \Eq{BBEuclidean} after the application of gradient flow, which also serves to renormalize all operators.
But the relation between the operator correlations and $\kappa_E,\kappa_B$ are known within the $\MSBAR$ scheme \cite{Laine:2021uzs}.
To relate the two procedures, we will compute the correlators in \Eq{continuation} and \Eq{BBEuclidean} at NLO in $\MSBAR$, both with gradient flow and without gradient flow.
To simplify the calculation we fix $\tau$ but take $\beta \to \infty$, that is, we compute vacuum correlators.
This is sufficient to establish the relation between the flowed and unflowed operator normalizations.
Gradient flow depends on a gradient-flow depth $\tauf$ which has units of distance-squared.
If $\sqrt{\tauf} \gtrsim \tau$ then flow effects significantly contaminate the intended correlator evaluation \cite{Eller:2018yje}, so in practice one is always interested in the $\tauf / \tau^2 \ll 1$ limit.
We will only attempt to compute correlation functions to lowest order in this limit (small gradient flow time expansion).
That is, we take the limit $\tauf \ll \tau^2 \ll \beta^2$.
To keep track of this limit we will introduce
$\chi \equiv \tau^2 / 8\tauf$ and we will be interested in $\chi \gg 1$.

To make the paper more self-contained, we will briefly review gradient flow.
Gauge field $A_\mu(x)$ correlations are evaluated within the path integral with $\MSBAR$ renormalization as usual.
Operators are constructed out of modified fields $B_\mu(x,\tauf)$
which are constructed from the gauge fields $A_\mu(x)$ via the following procedure.
One introduces a fictitious gradient-flow time $\tauf$, which determines a level of UV-smoothing of the gauge fields.
Specifically, at $\tauf=0$ the gauge fields contain fluctuations at all scales ($B_\mu(x,0) = A_\mu$ the fields appearing in the path integral), and at subsequent $\tauf$ values they are defined by the differential equation
\cite{luscher_properties_2010} 
\begin{align}
\label{defineGradientFlow}
\frac{d}{d\tauf}B_{\nu}(\tauf) & = D_{\mu}G_{\mu\nu}
+ \xi D_\nu \partial_\mu B_\mu \,,
\nonumber \\ 
G_{\mu\nu} & \equiv \partial_{\mu}B_{\nu}-\partial_{\nu}B_{\mu}+\left[B_{\mu},B_{\nu}\right] \\ \nonumber
D_{\mu} & \equiv \partial_{\mu}+\left[B_{\mu}, \;\right].
\end{align}
\begin{figure}[b]
\centering{}
\begin{tabular}{ccc}
\includegraphics[scale=0.11]{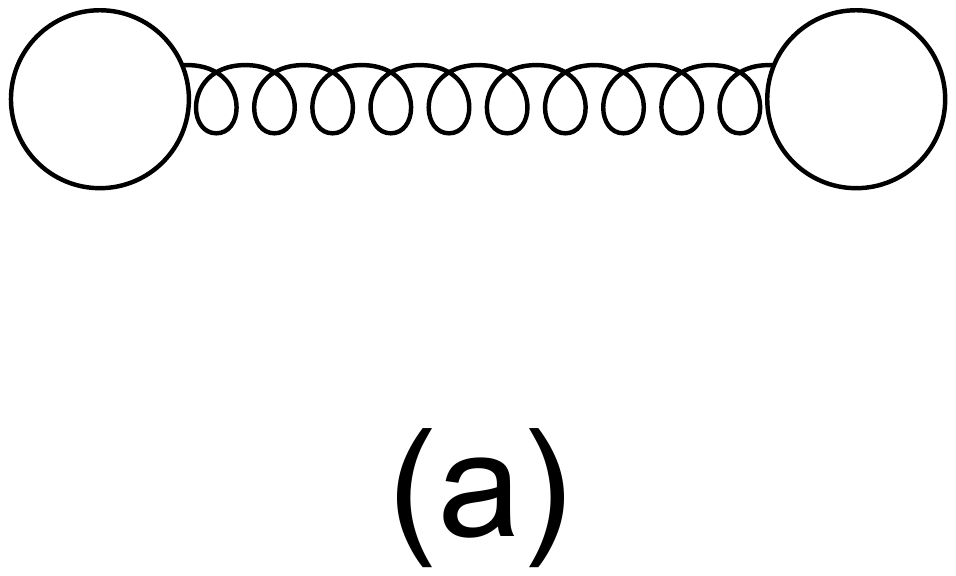} & \includegraphics[scale=0.11]{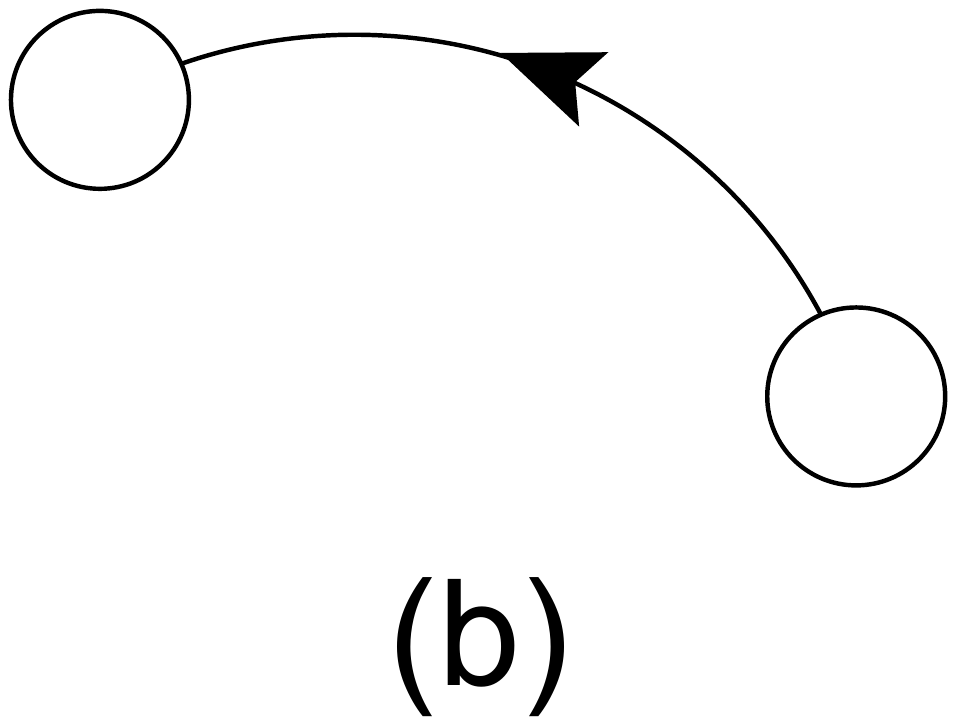} & \includegraphics[scale=0.11]{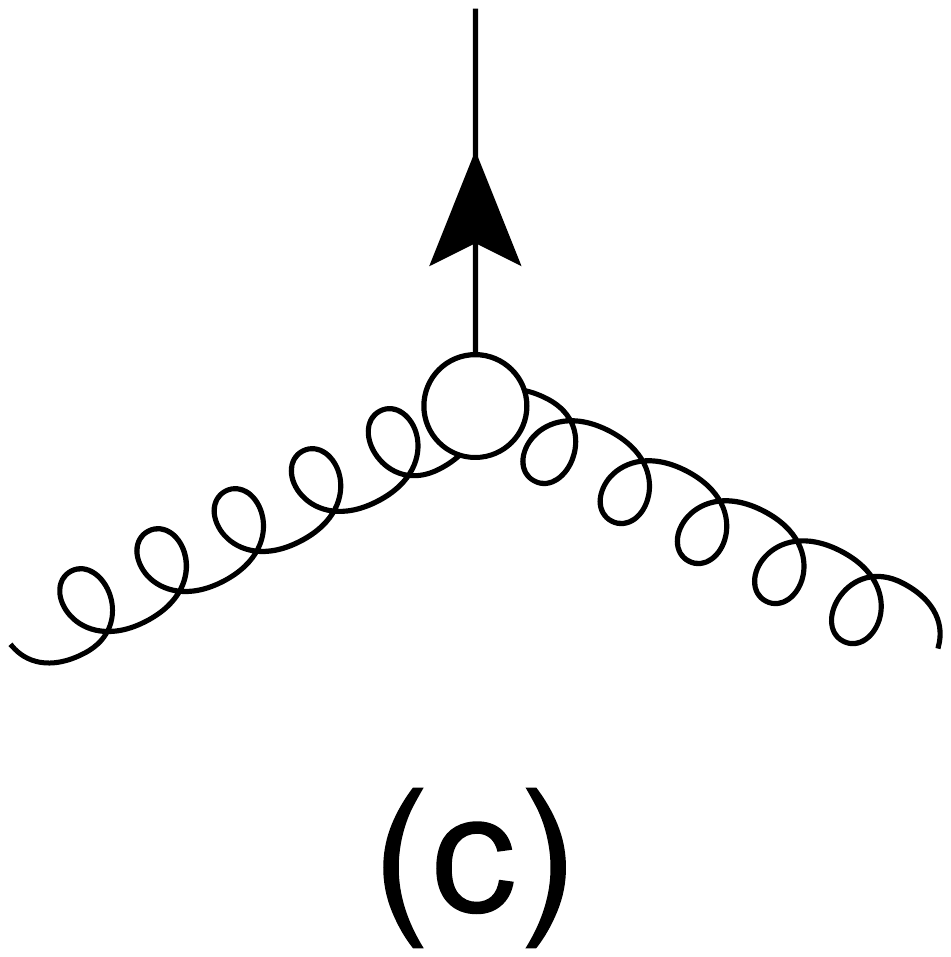} \vspace{-5ex} \tabularnewline  \includegraphics[scale=0.11]{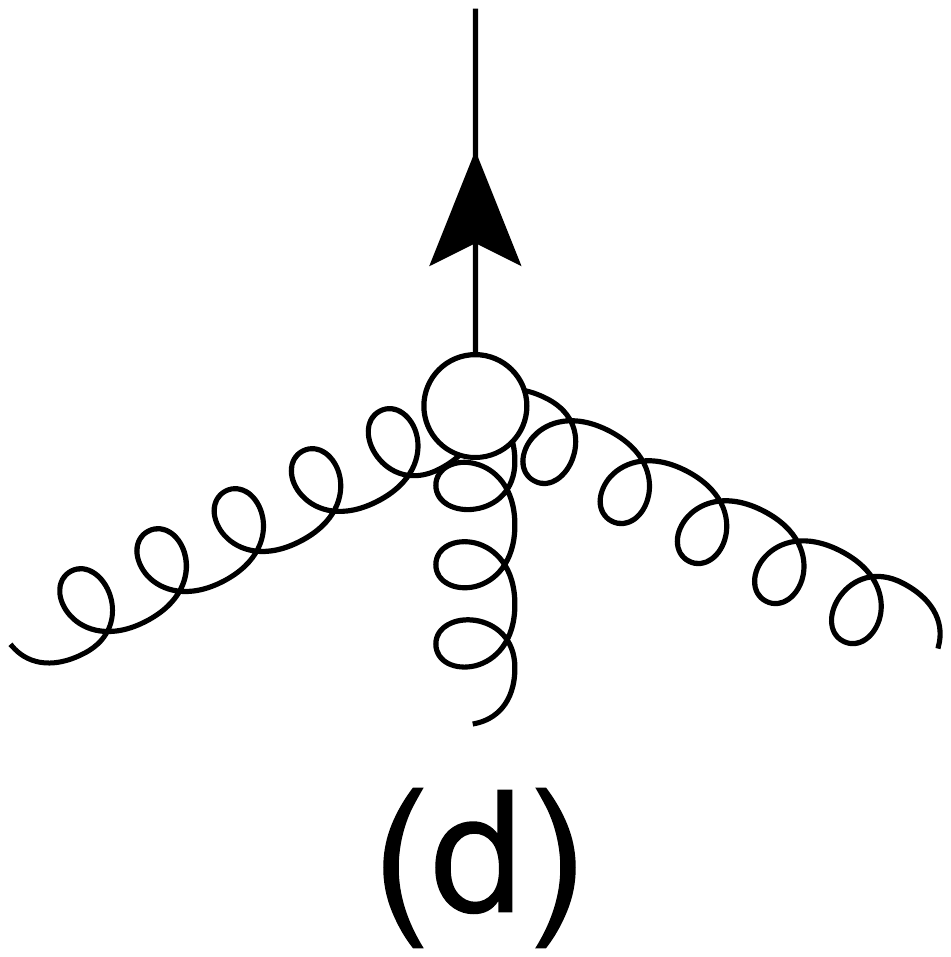} & \includegraphics[scale=0.11]{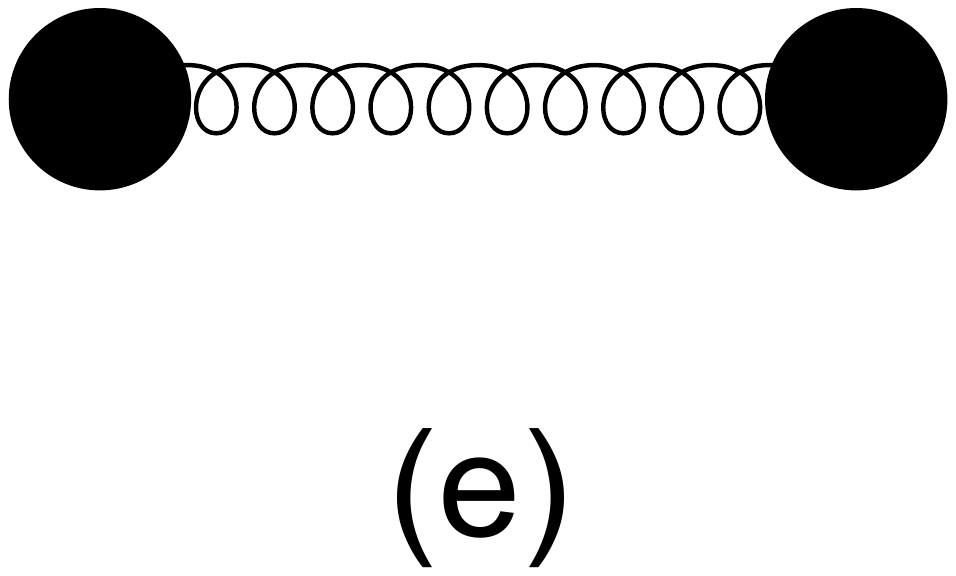} & \includegraphics[scale=0.11]{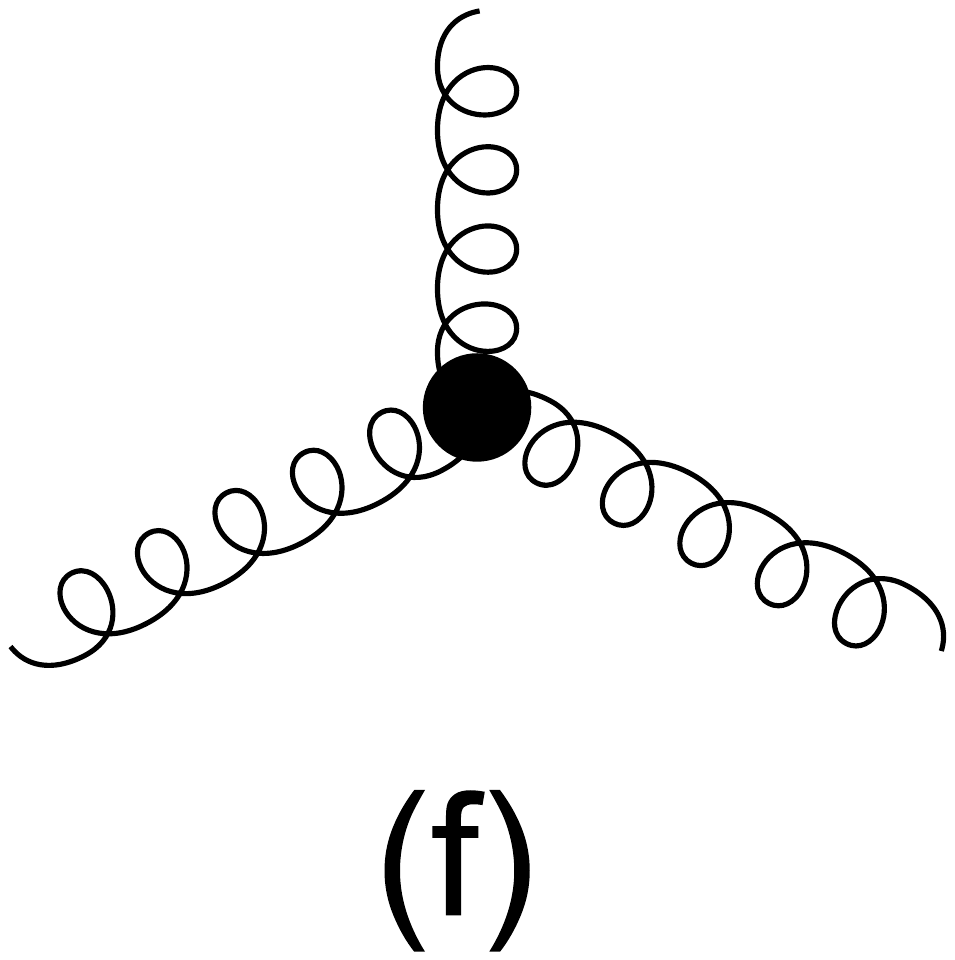} \vspace{-5ex} \tabularnewline  \includegraphics[scale=0.11]{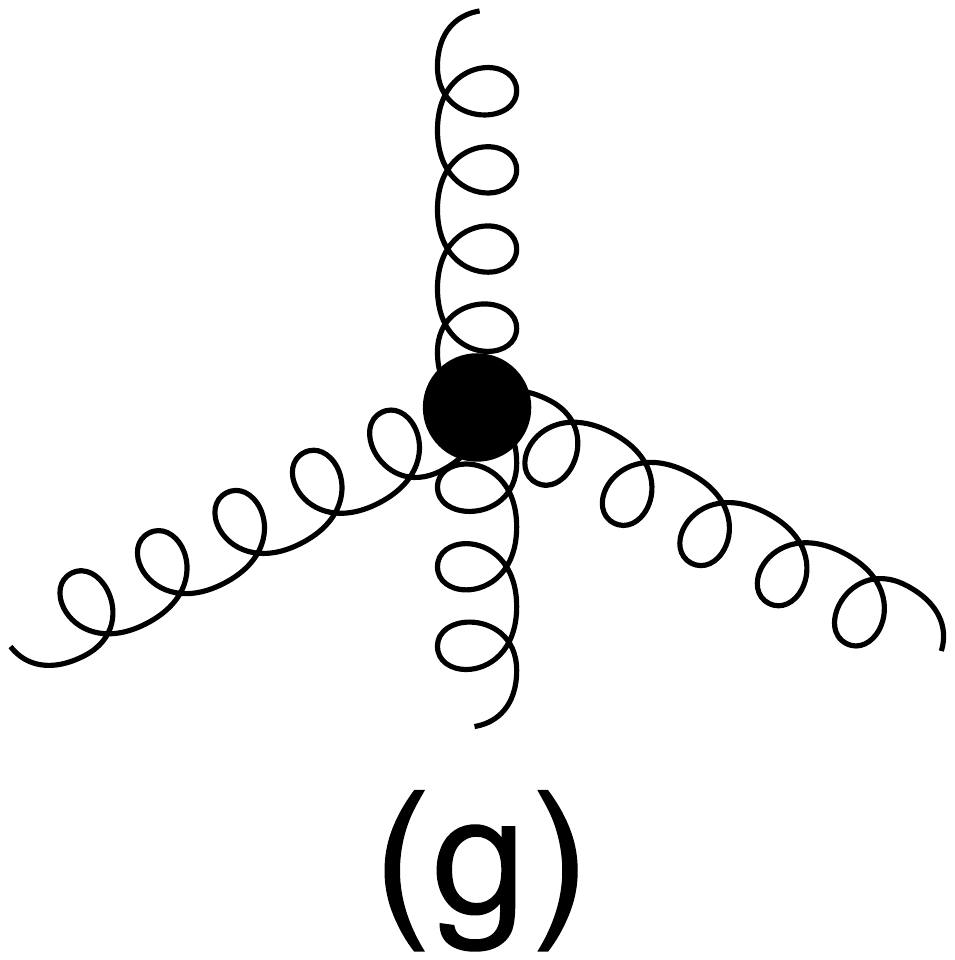} & \includegraphics[scale=0.11]{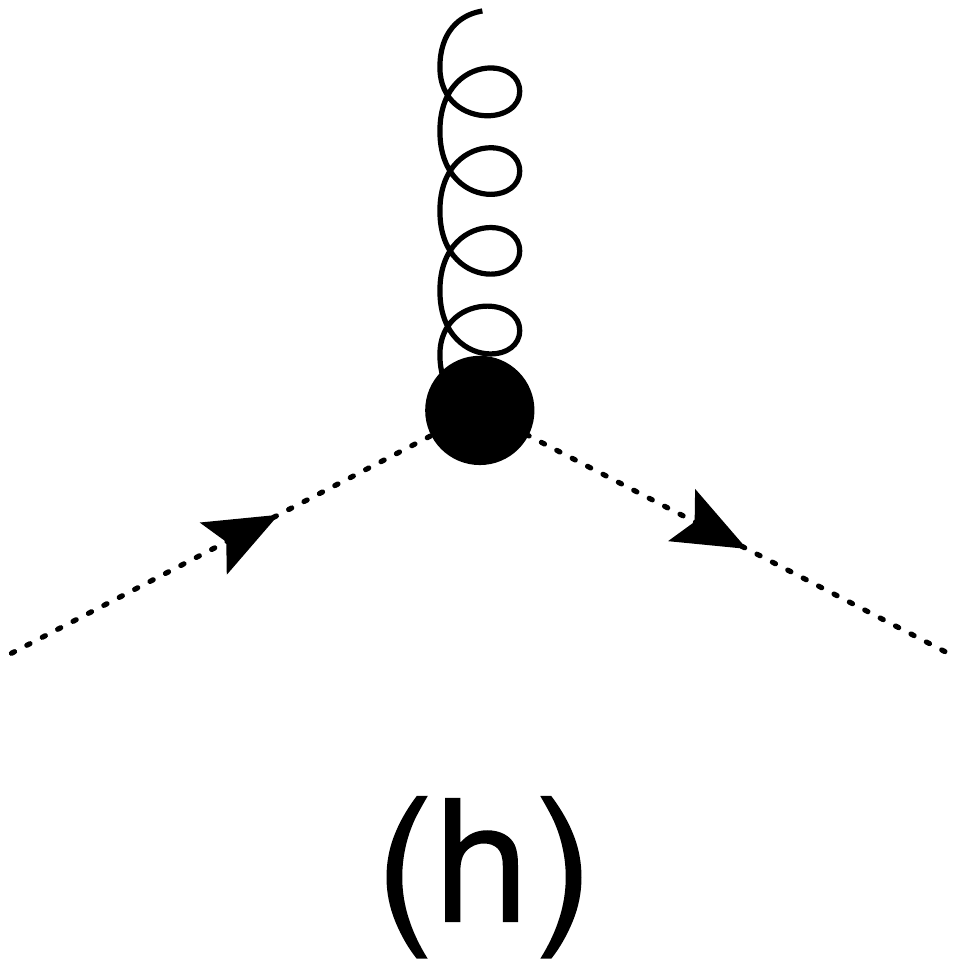} & \includegraphics[scale=0.11]{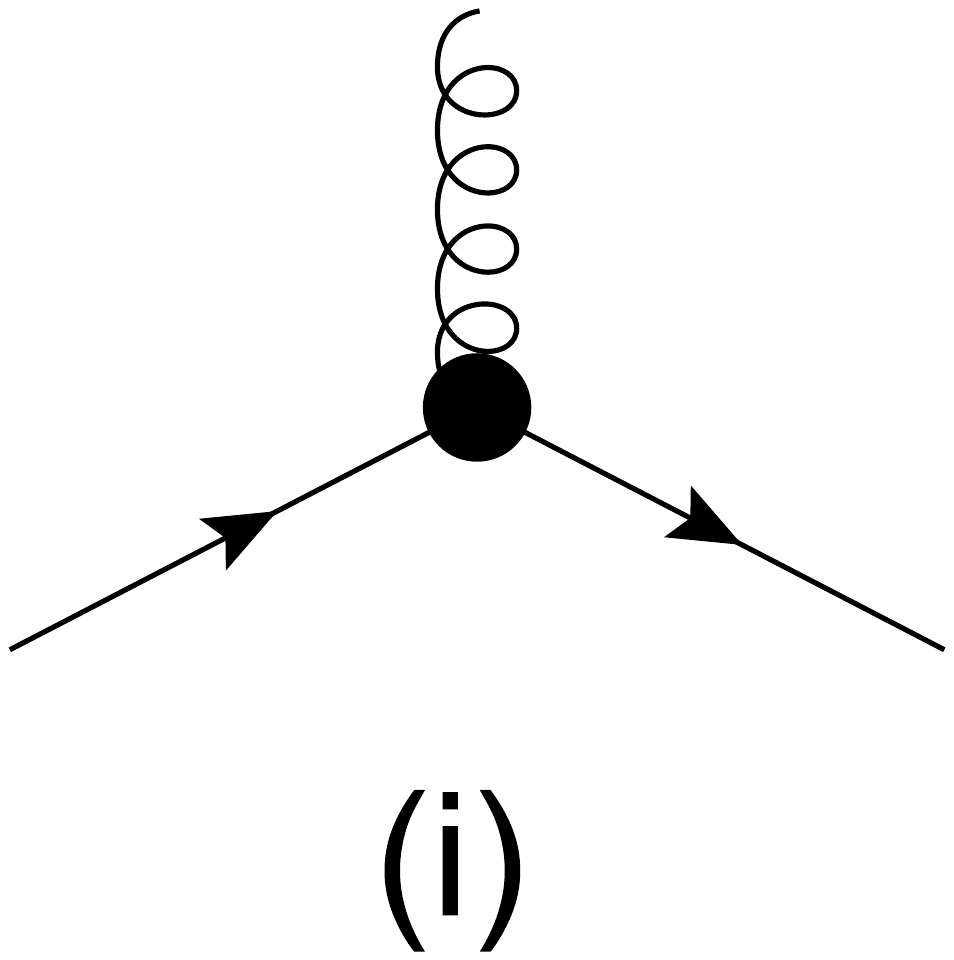} \vspace{-6ex}
\end{tabular}\caption{Feynman diagram components.
$(a)$ represents the correlator of two flowed ($B$) fields, $(b)$ is the $BL$ correlator which occurs purely during the gradient flow, $(c)$ and $(d)$ are the 3-point and 4-point vertices which arise during gradient flow, and the remaining elements are the standard objects in unflowed perturbation theory.}
\label{Un/flowed Feynman rules}
\end{figure}
\begin{figure}[t]
\centering{}
\begin{tabular}{cc}
\vspace{-2ex}
\includegraphics[scale=0.11]{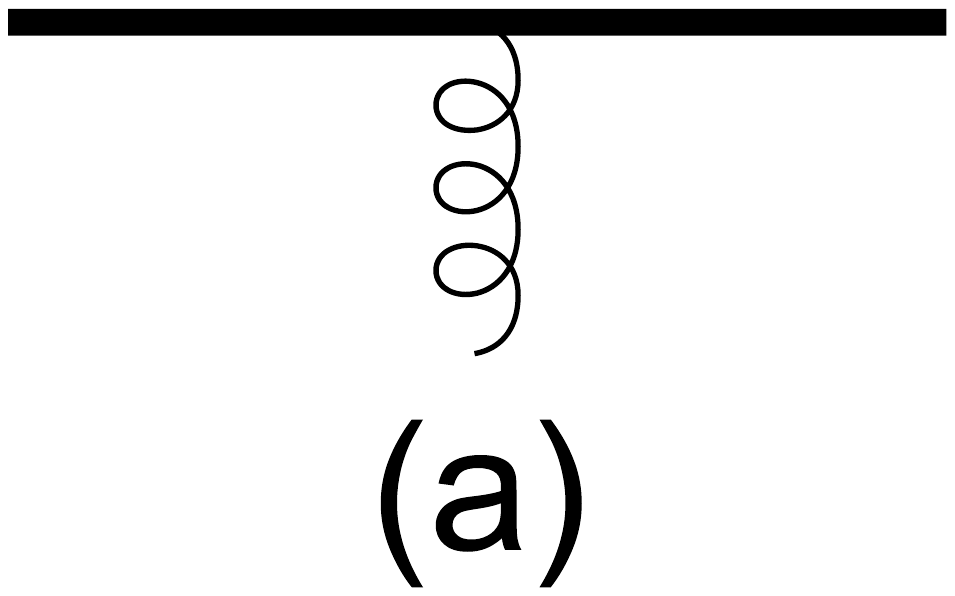} &
\includegraphics[scale=0.11]{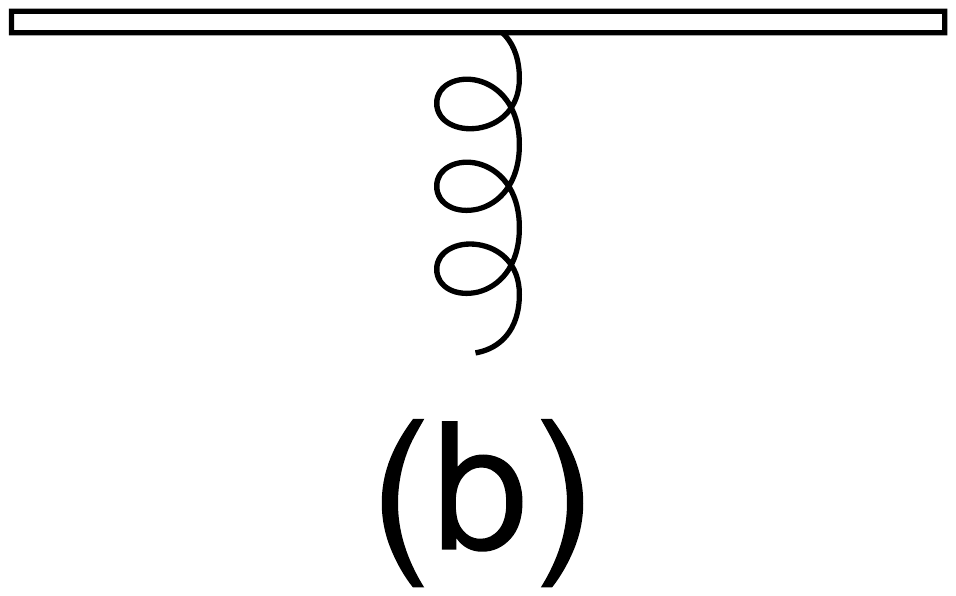} \vspace{-6ex} 
\end{tabular}
\caption{Vertices associated with an unflowed (left) or flowed (right) Wilson line.
The transverse coordinates of the vertex are fixed but the longitudinal coordinate should be integrated along the available length of the Wilson line.}
\label{fig:FeynmanWilson}
\end{figure}
The second $\propto \xi$ term in the first line evolves the gauge fixing choice with gradient flow depth, and all results should be independent of $\xi$.
At the linearized level for the choice $\xi=1$ (analogous to Feynman gauge) we have
$B_\mu(p,\tauf) = A_\mu(p) \,\exp(-\tauf p^2)$
but the nonlinearities in the evolution equations introduce additional vertices which make the Feynman rules slightly more complicated than in the unflowed theory.
Specifically, there are two sorts of vertices, those which occur between $A_\mu$ fields in the path integral and those which arise from the nonlinearities of gradient flow.
And there are two kinds of propagators:  those between either $A_\mu$ fields or $B_\mu$ fields via their lowest-order relation to an $A_\mu$ field,
and those which occur purely within the gradient-flow process, between a $B_\mu$ field and an auxiliary $L$-field.
For simplicity, this paper will consider only $\xi = 1$ both during flow and in the unflowed path integral (Feynman and Feynman-flow gauge).
We indicate the Wilson line as a double solid line and the $E,B$ field insertions on it as open blobs.
Otherwise our notation follows L\"uscher
\cite{luscher_perturbative_2011}:
Finite flow-time vertices are open and path-integral vertices are solid blobs, and propagators which exist only at finite flow time (BL propagators) are lines with arrows indicating the direction towards larger flow time.
The objects appearing in the Feynman diagrams are shown in Figure \ref{Un/flowed Feynman rules},
the additional objects arising from the Wilson lines are shown in Figure \ref{fig:FeynmanWilson}, and the detailed rules are presented in Appendix \ref{secFeynRules}.

\section{Calculational details}
\label{sec:diagrams}

\begin{figure}[htb]
\centering{}
\begin{tabular}{cccc}
\includegraphics[scale=0.1]{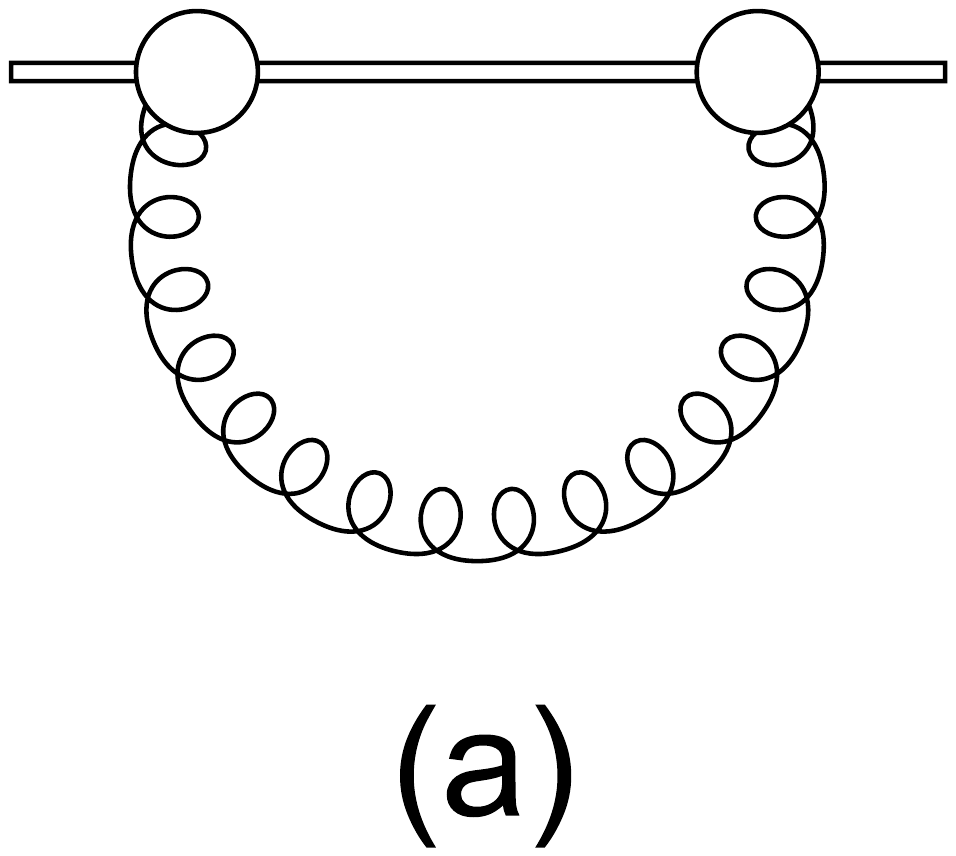} \hspace{-1.5em} & \includegraphics[scale=0.1]{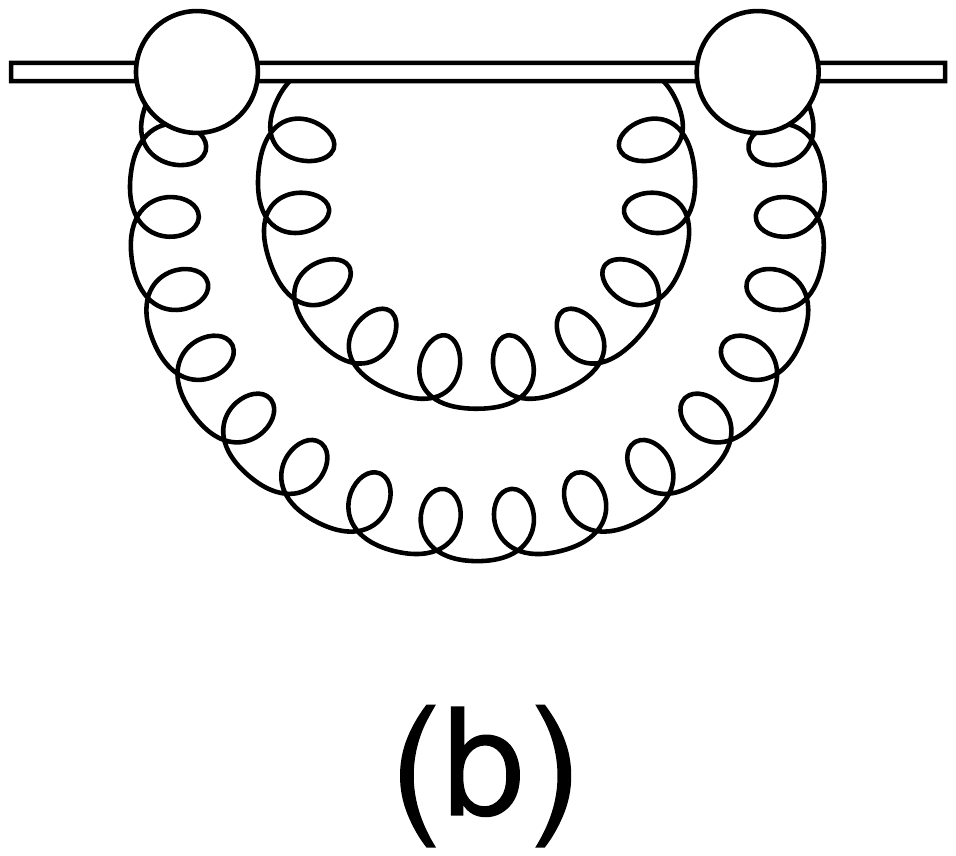} \hspace{-1.5em} & \includegraphics[scale=0.1]{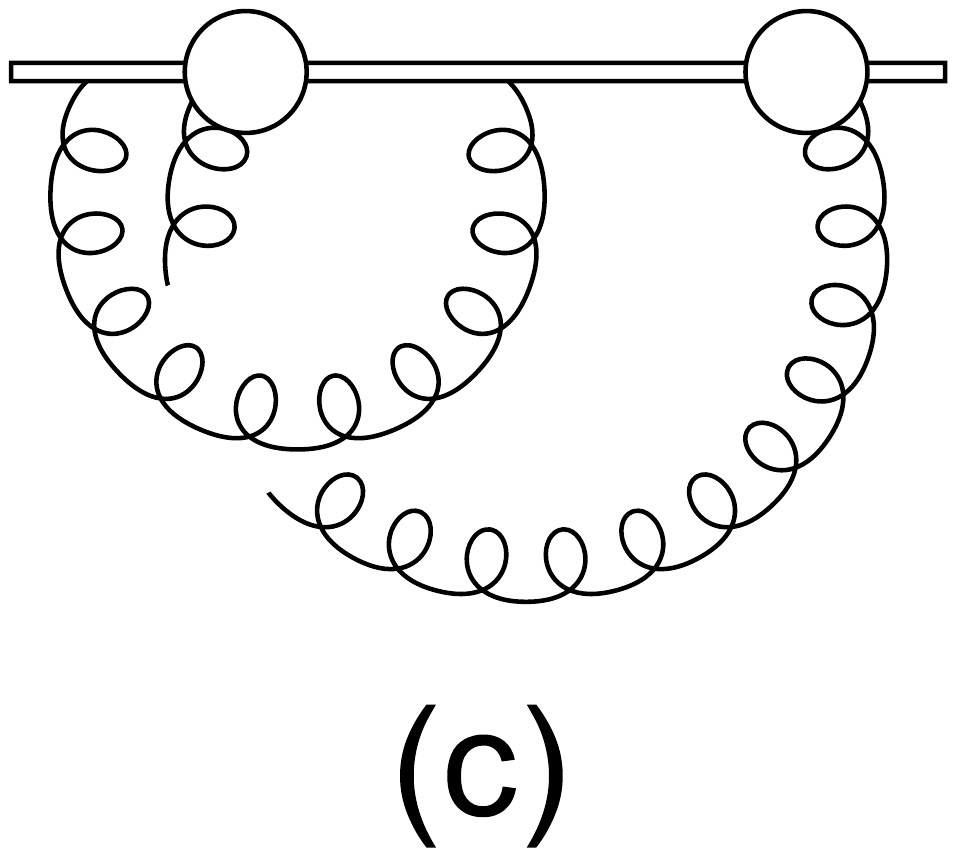} \hspace{-1.5em} & \includegraphics[scale=0.1]{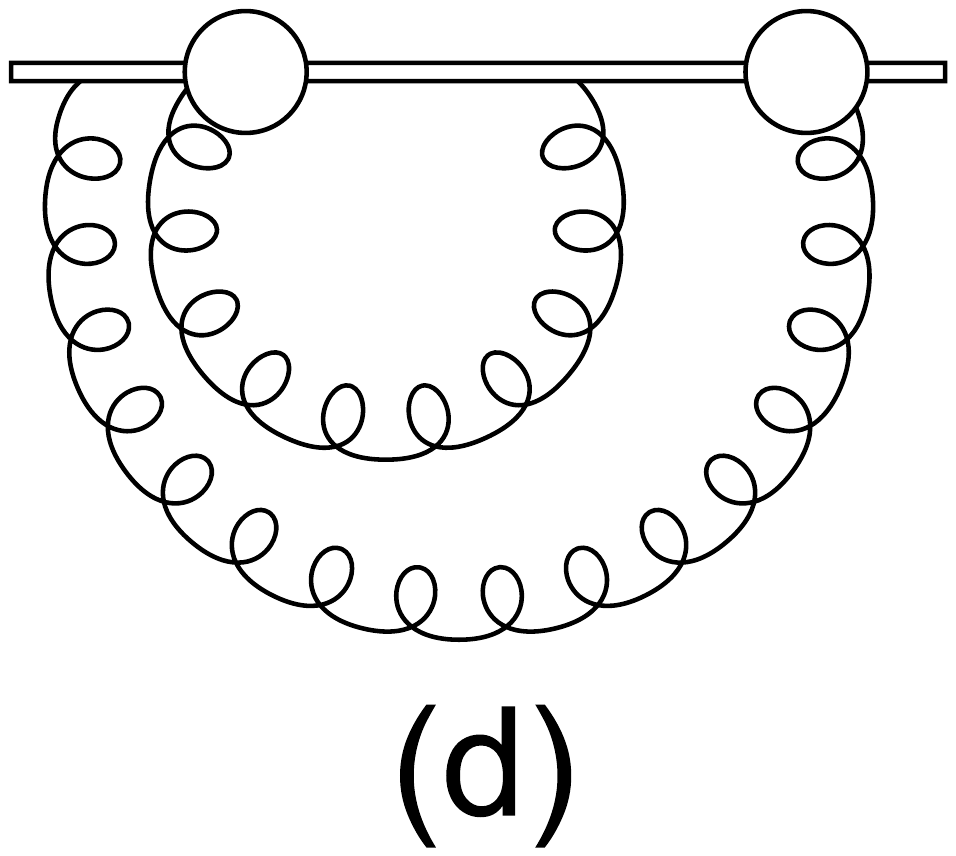}\vspace{-5ex} \tabularnewline \includegraphics[scale=0.1]{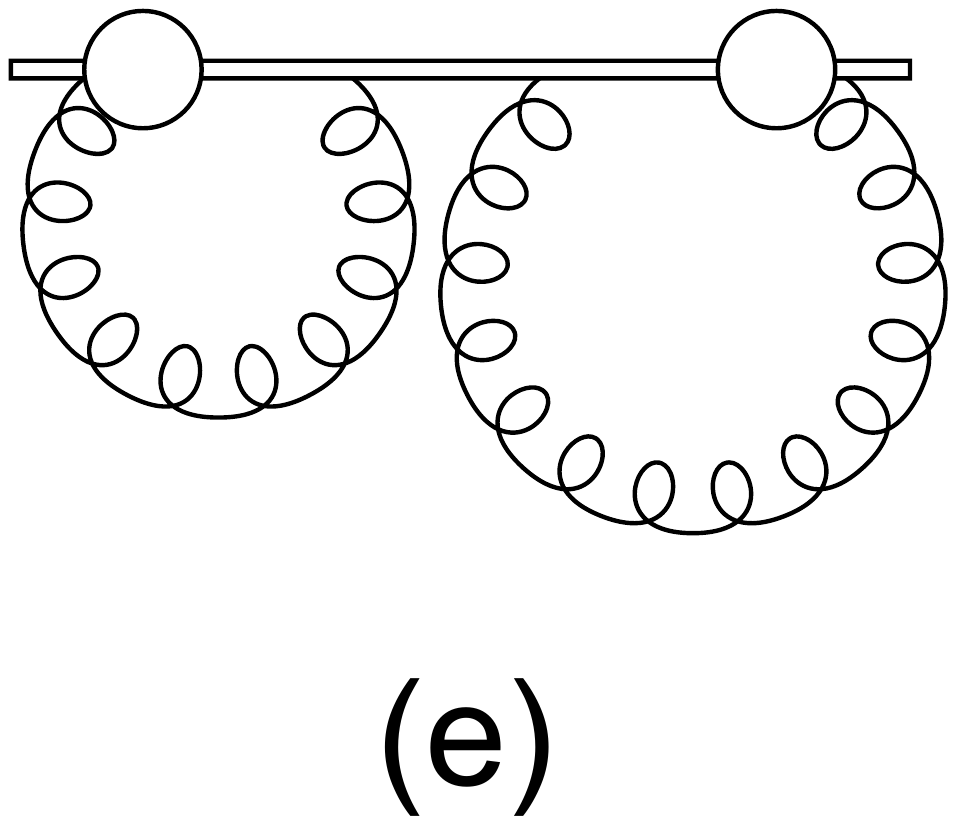} \hspace{-1.5em} & \includegraphics[scale=0.1]{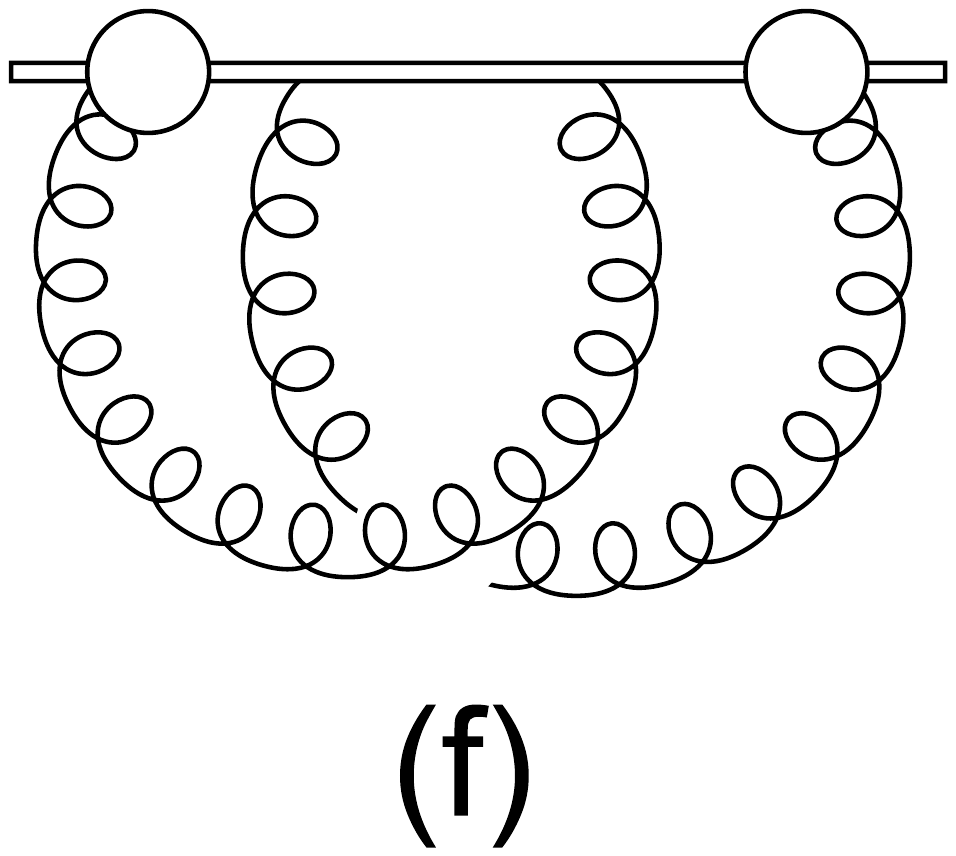} \hspace{-1.5em} &  \includegraphics[scale=0.1]{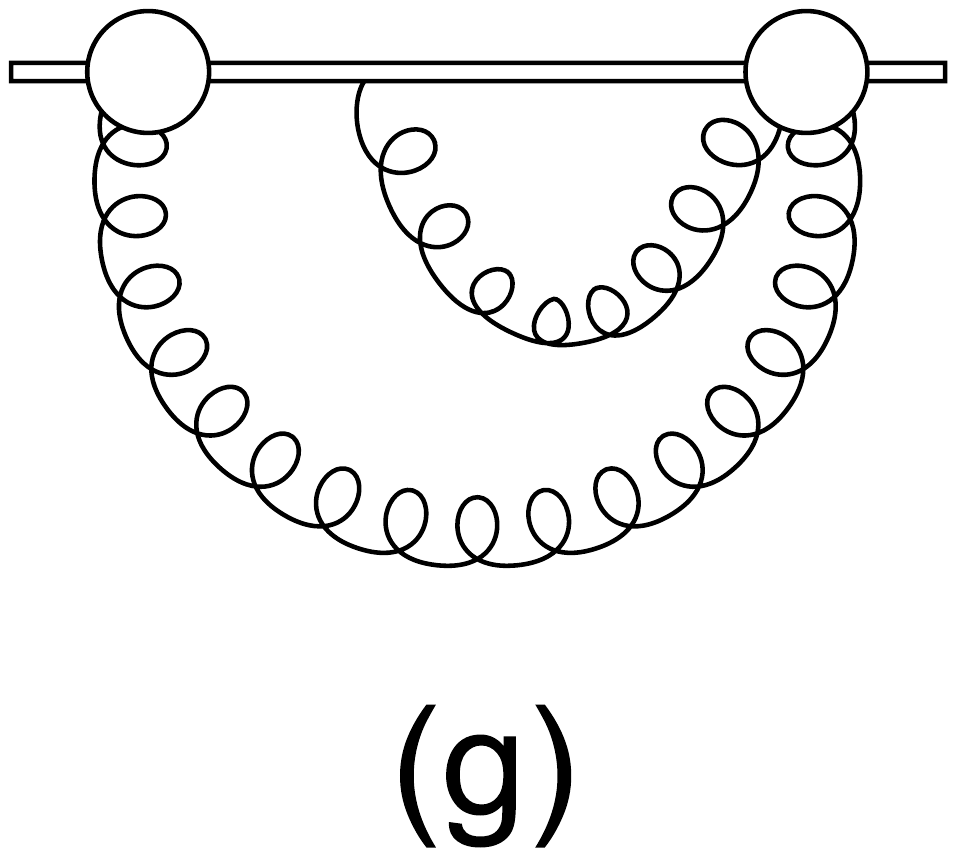} \hspace{-1.5em} & \includegraphics[scale=0.1]{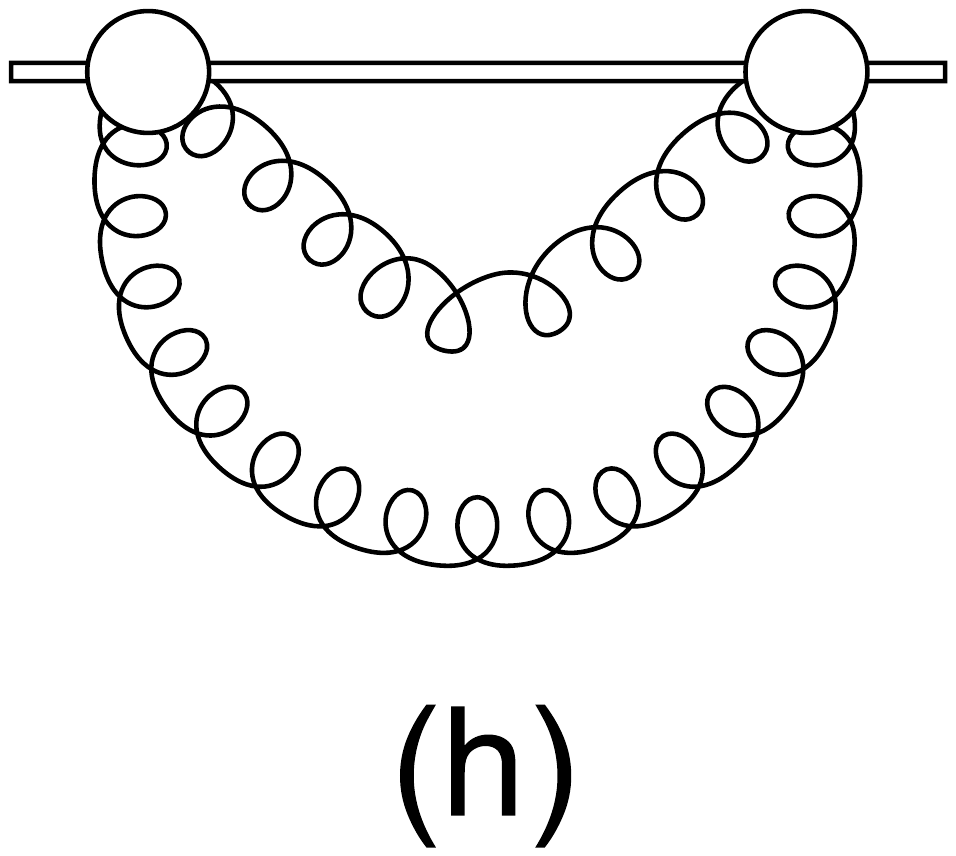}\vspace{-5ex} \tabularnewline \includegraphics[scale=0.1]{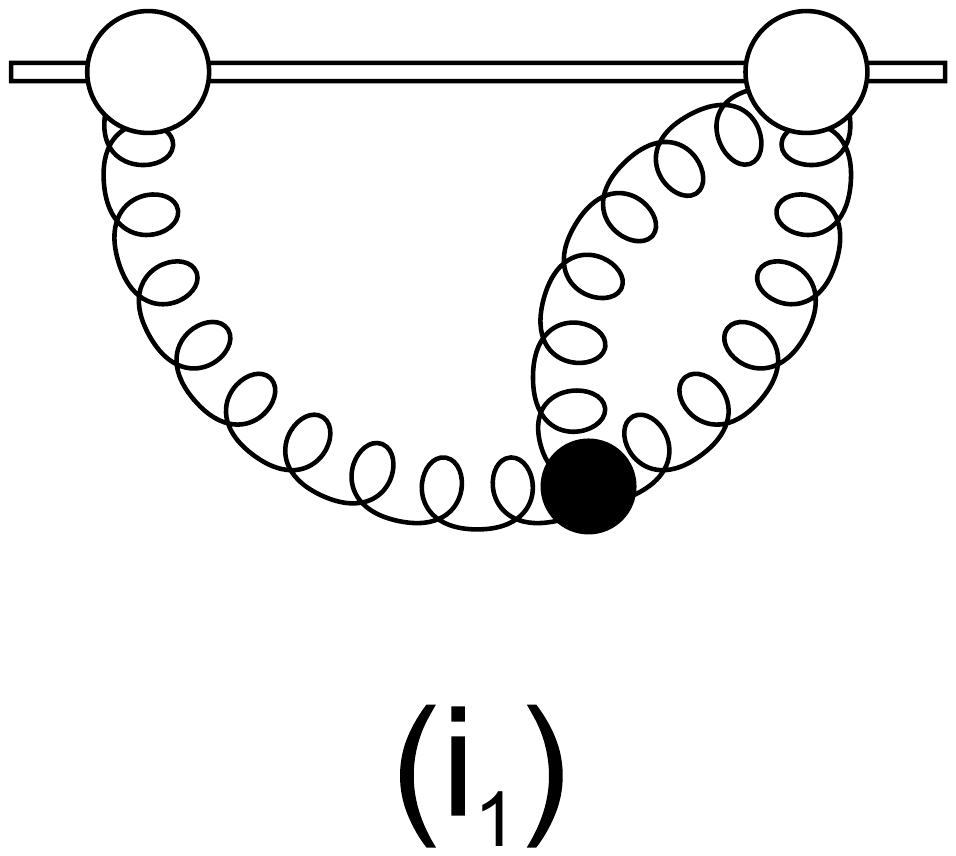} \hspace{-1.5em} & \includegraphics[scale=0.1]{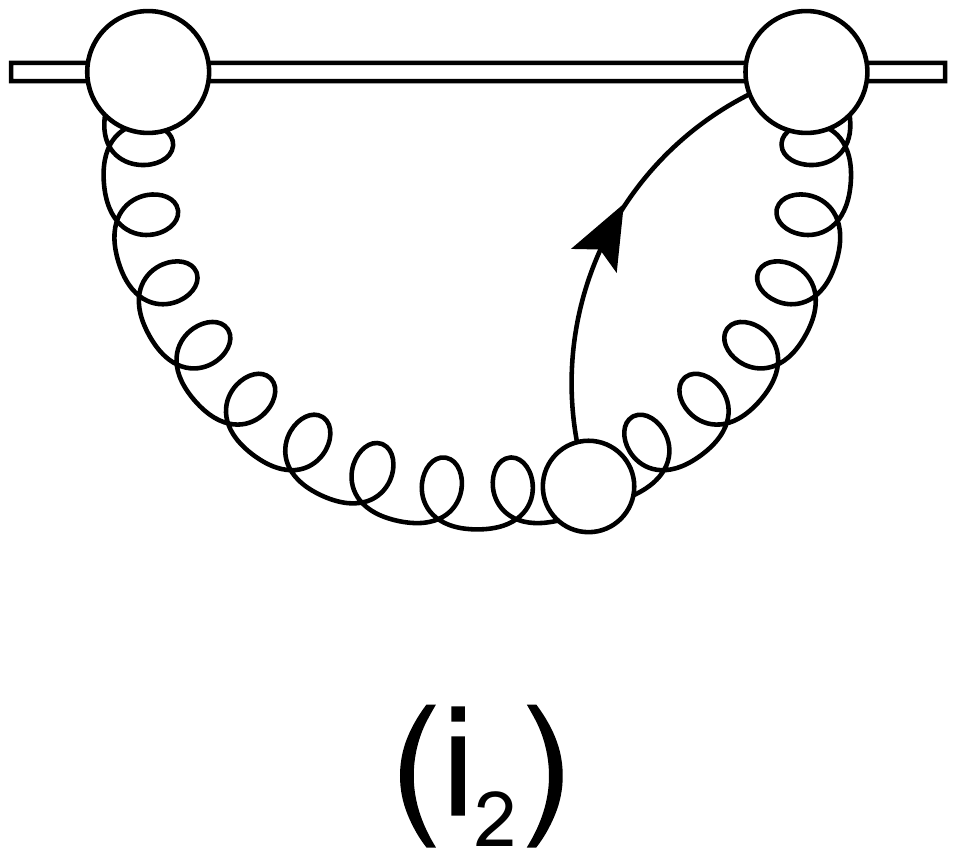} \hspace{-1.5em} &  \includegraphics[scale=0.1]{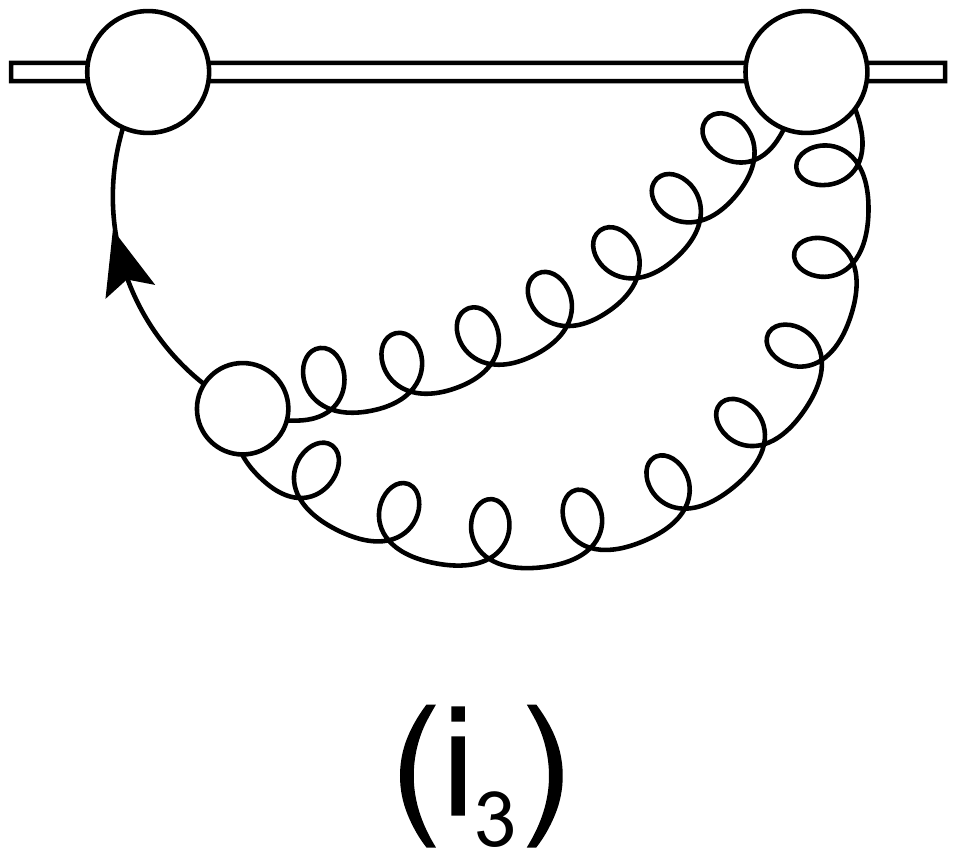} \hspace{-1.5em} & \includegraphics[scale=0.1]{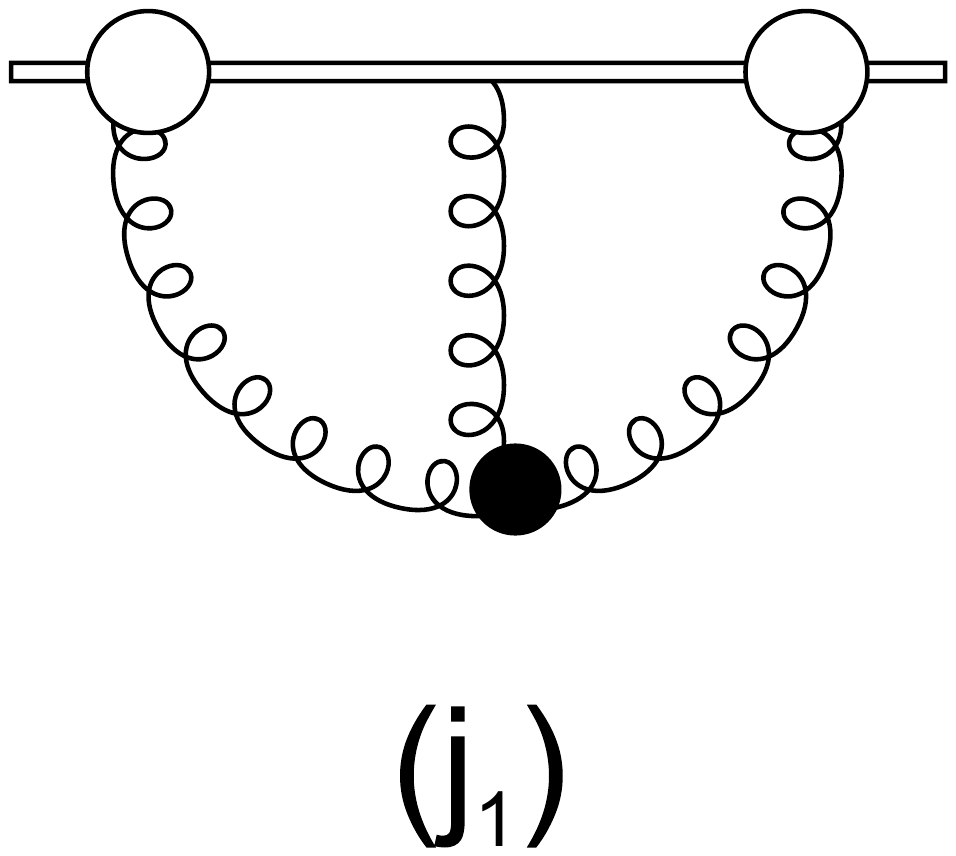}\vspace{-5ex} \tabularnewline \includegraphics[scale=0.1]{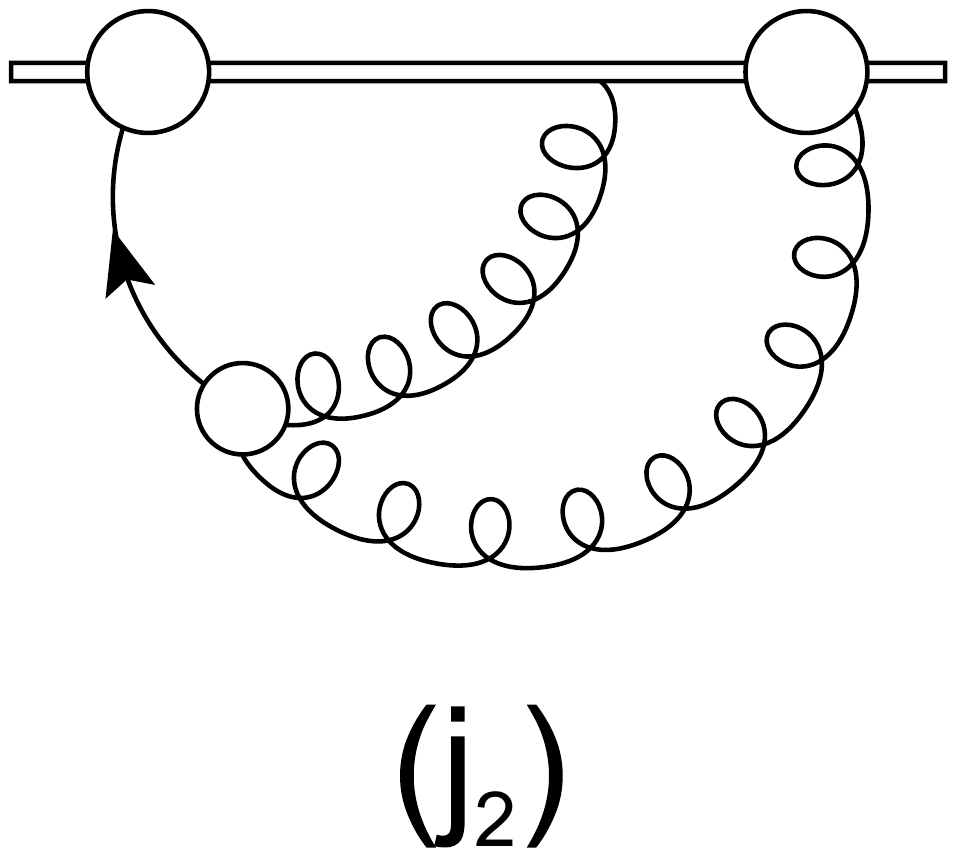} \hspace{-1.5em} & \includegraphics[scale=0.1]{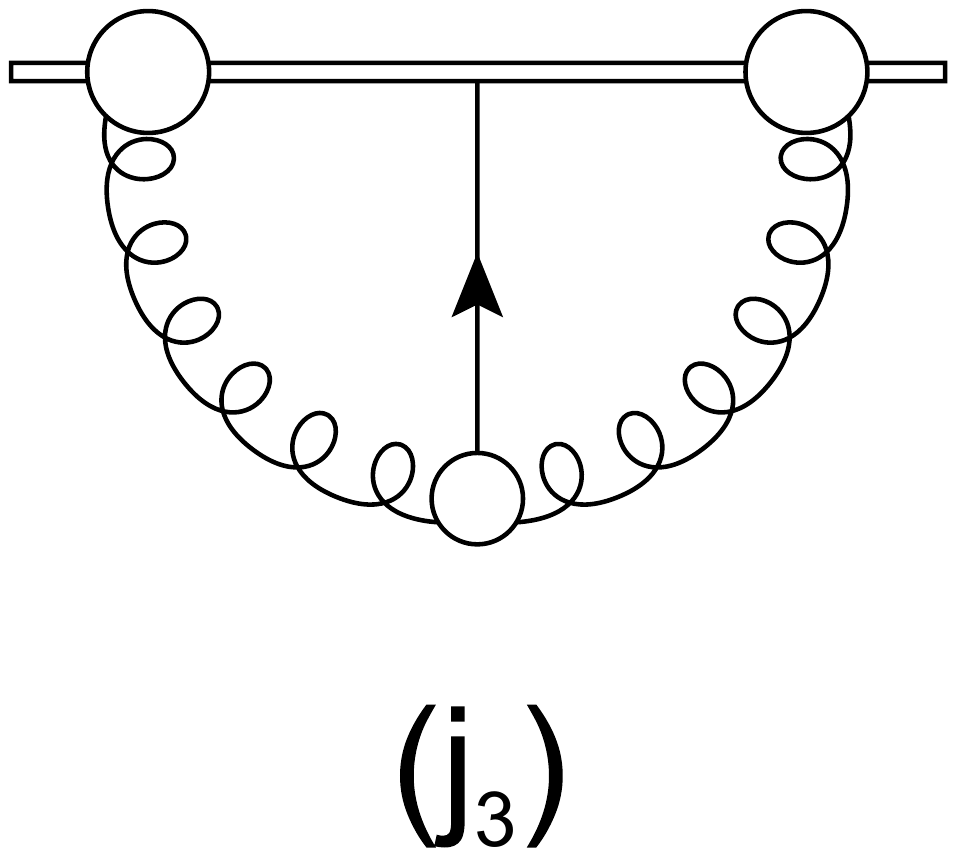} \hspace{-1.5em} & \includegraphics[scale=0.1]{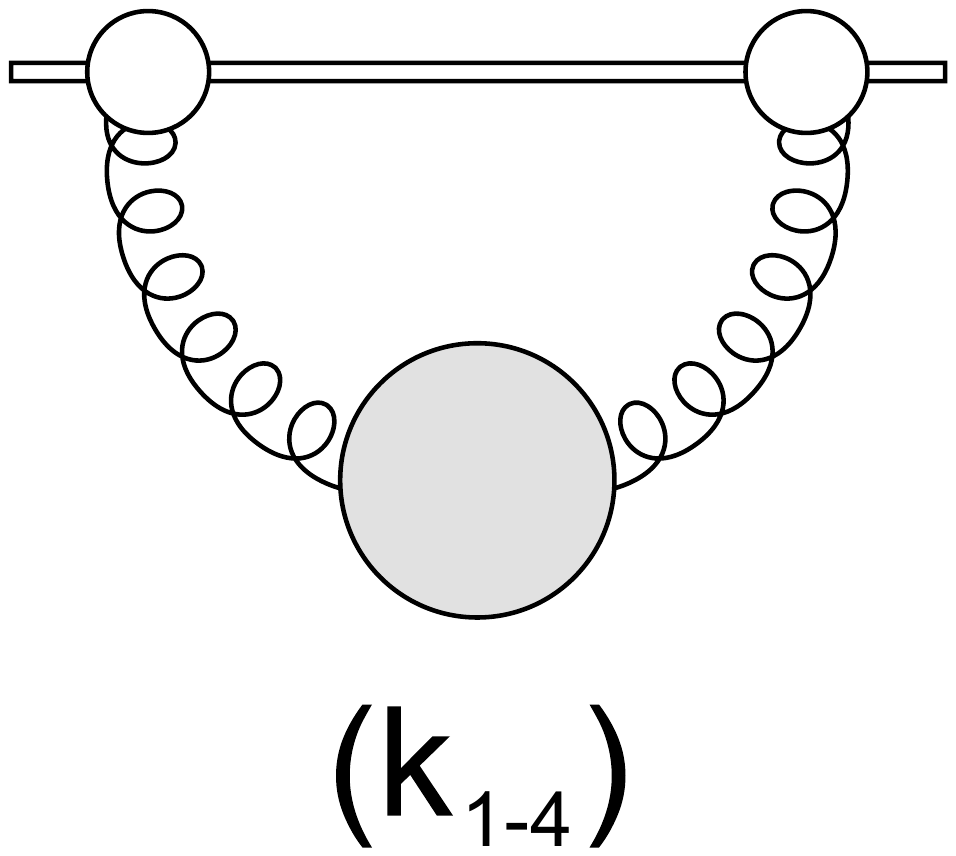} \hspace{-1.5em} & \includegraphics[scale=0.1]{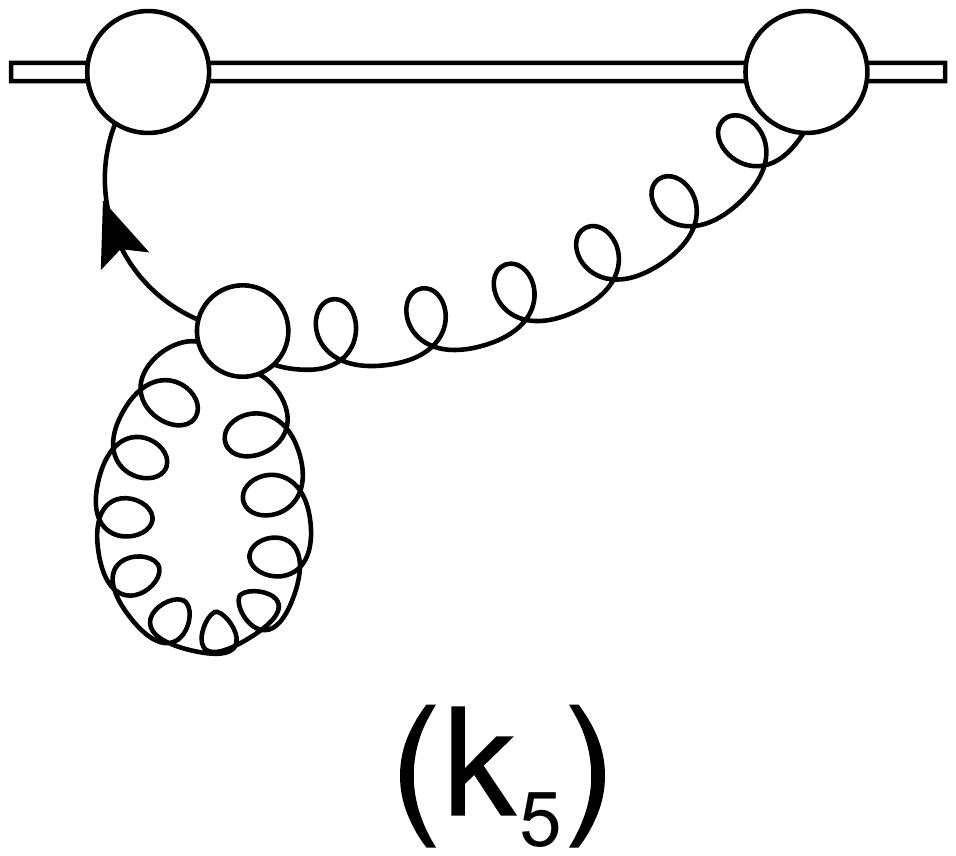}\vspace{-5ex} \tabularnewline \includegraphics[scale=0.1]{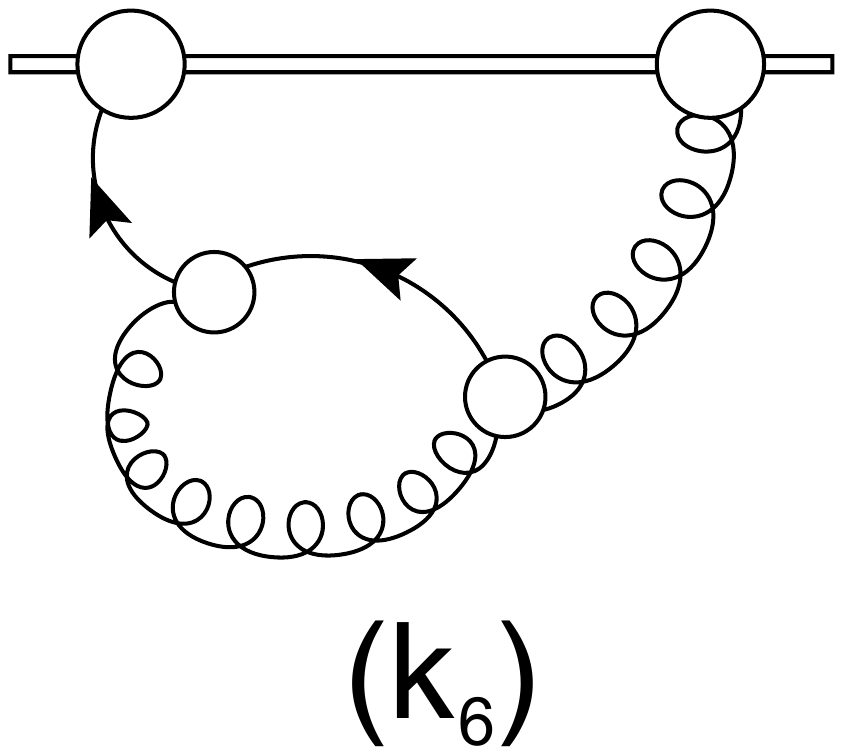} \hspace{-1.5em} & \includegraphics[scale=0.1]{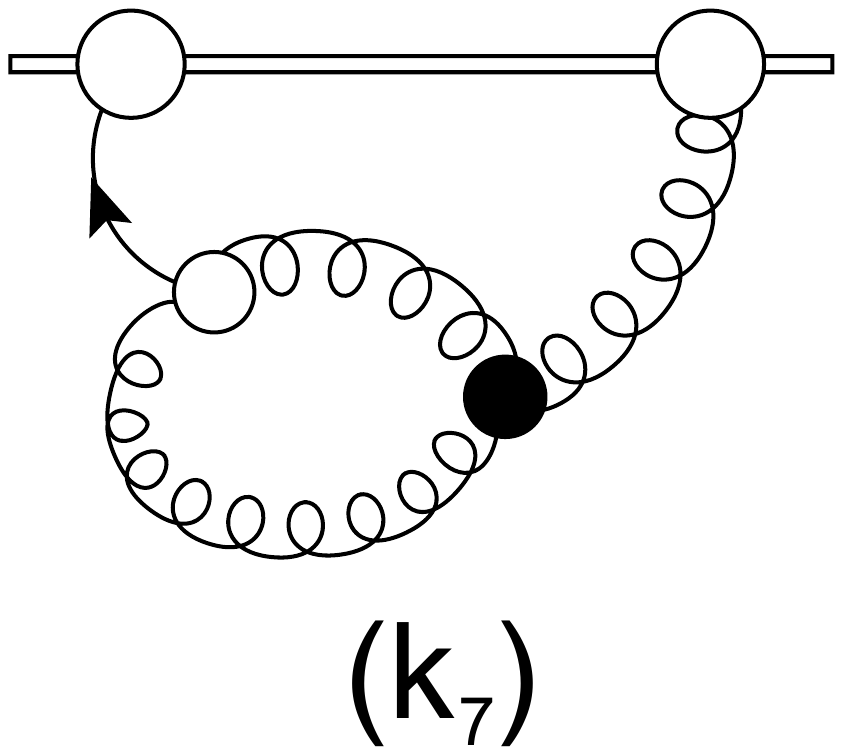} \hspace{-1.5em} & \includegraphics[scale=0.1]{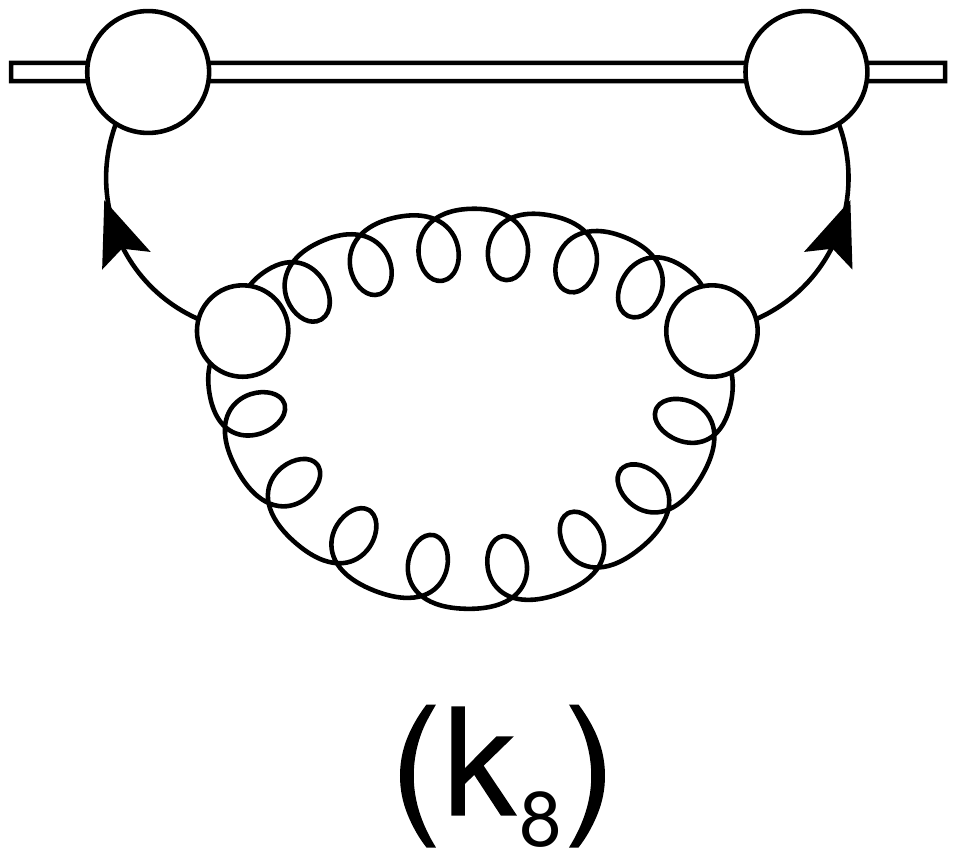} \vspace{-7ex}
\end{tabular}\caption{Diagrams with gradient flow, following
L\"uscher's notation \cite{luscher_perturbative_2011} and Burnier \textsl{et al} nomenclature \cite{Burnier:2010rp}.
The field strength is represented by a big white dot.
Small white dots are flowed vertices and black dots are unflowed vertices.
The arrows represent BL propagators.
The grey blob denotes the 1-loop unflowed self-energy.
Curvy lines are gluon propagators and the flowed Wilson line is a straight double line.}
\label{Flowed diagrams}
\end{figure}

\begin{figure}
\centering{}
\begin{tabular}{cccc}

\includegraphics[scale=0.1]{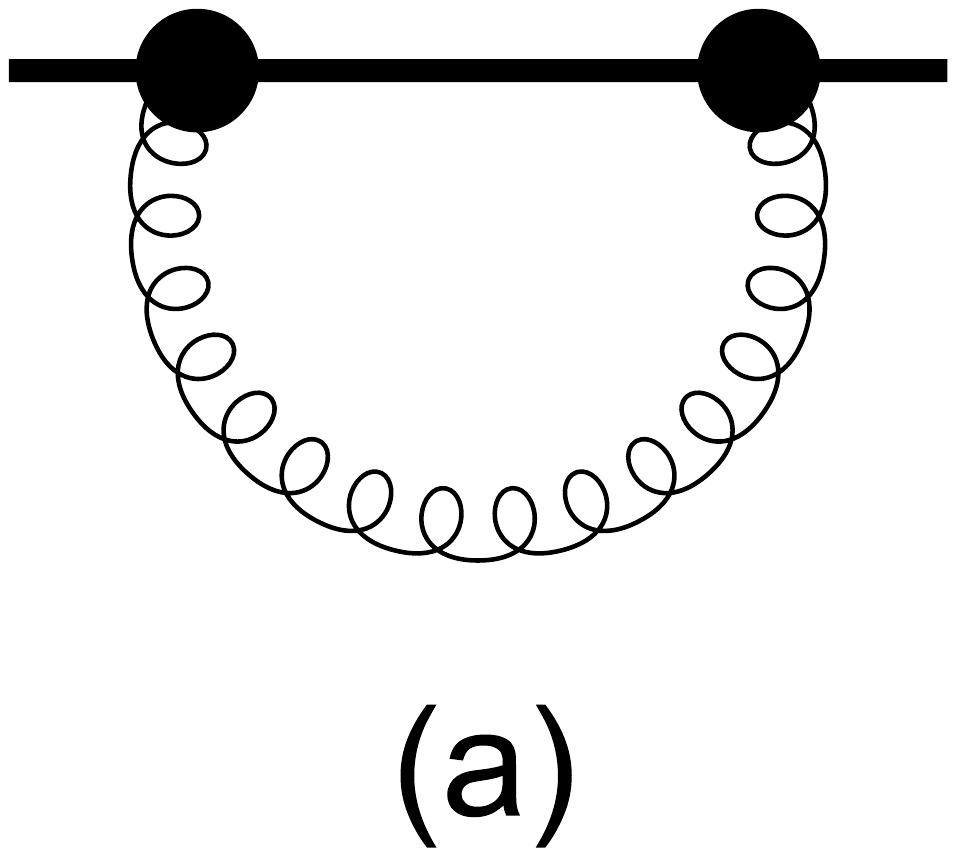} \hspace{-1em} & \includegraphics[scale=0.1]{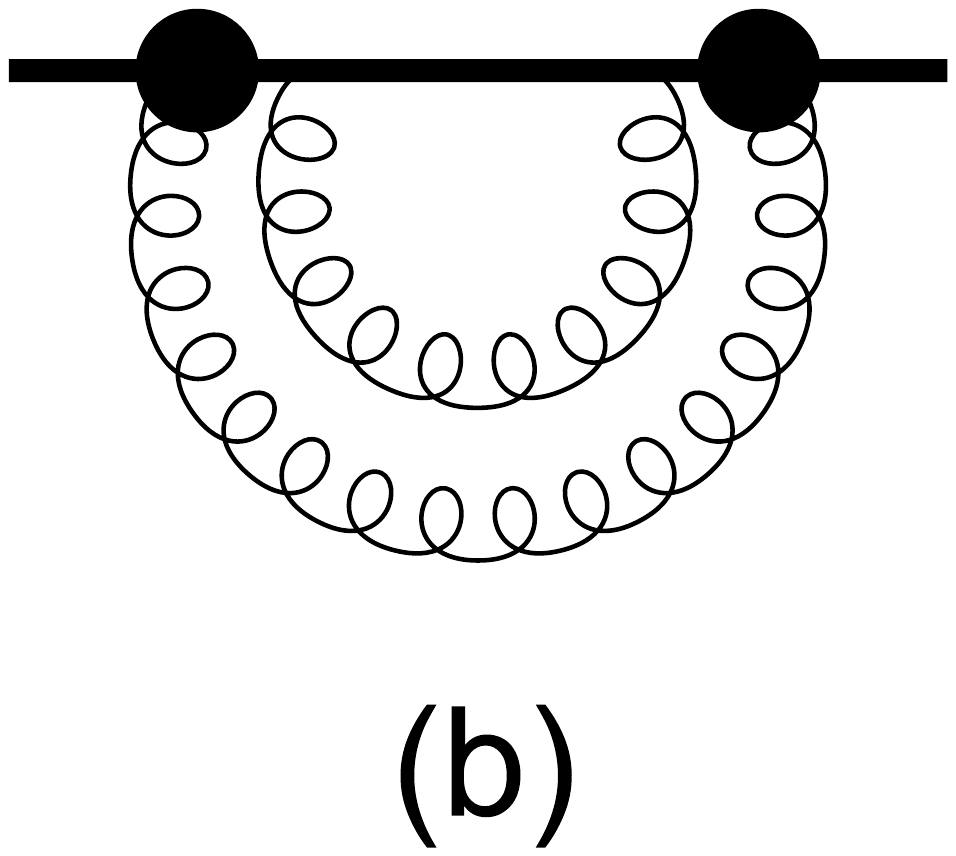} \hspace{-1.5em} & \includegraphics[scale=0.1]{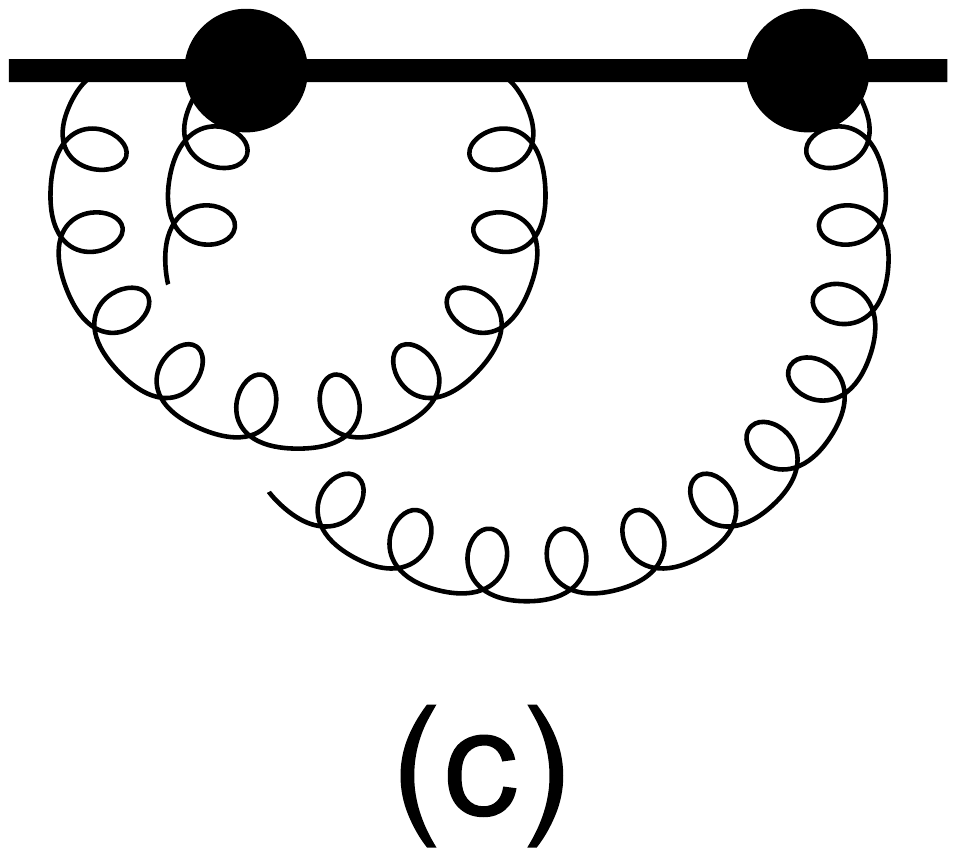} \hspace{-1.5em} & \includegraphics[scale=0.1]{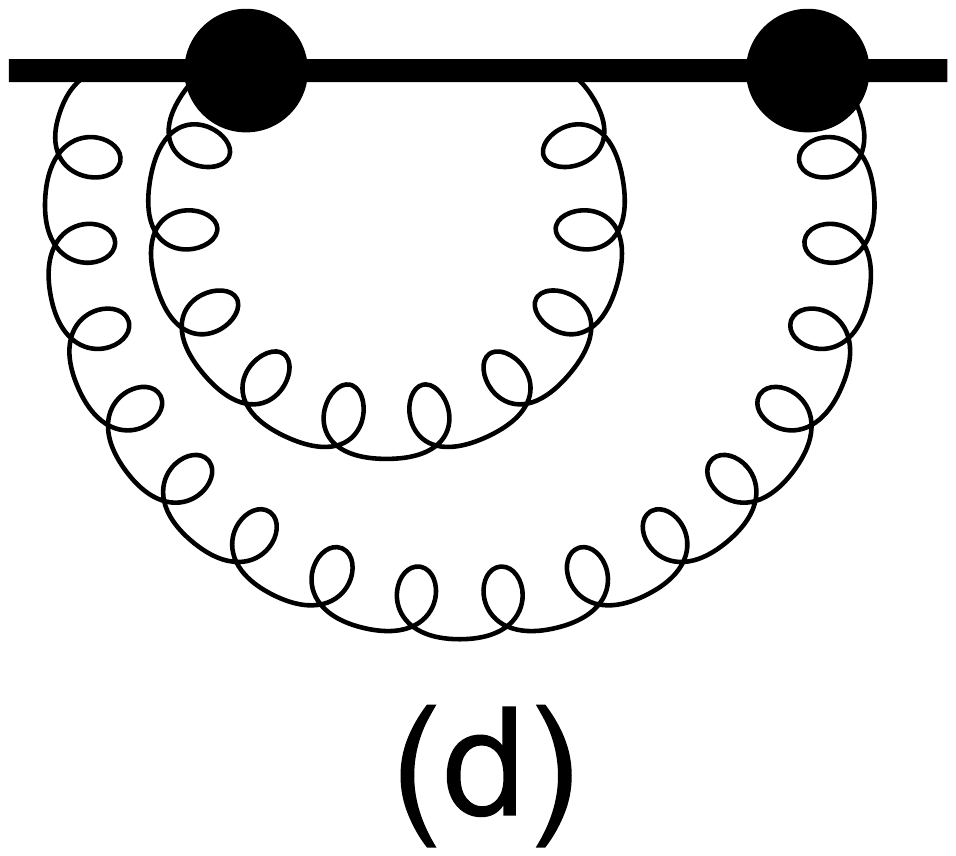}\vspace{-5ex} \tabularnewline  \includegraphics[scale=0.1]{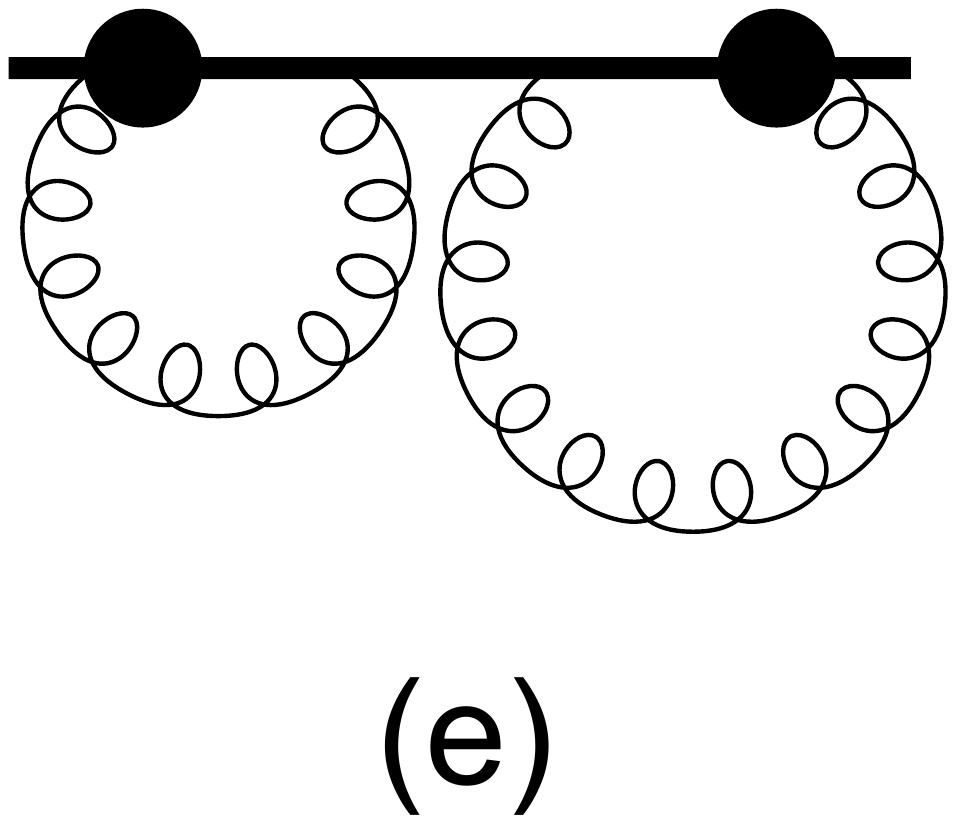} \hspace{-1em} & \includegraphics[scale=0.1]{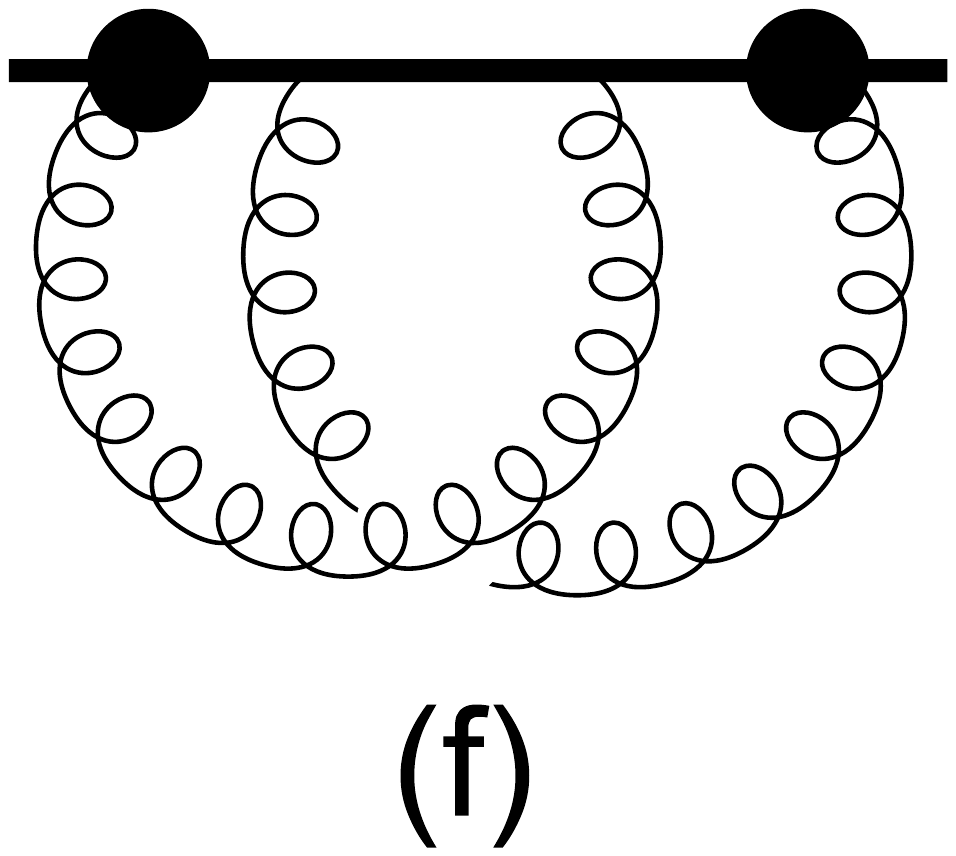} \hspace{-1.5em} &  \includegraphics[scale=0.1]{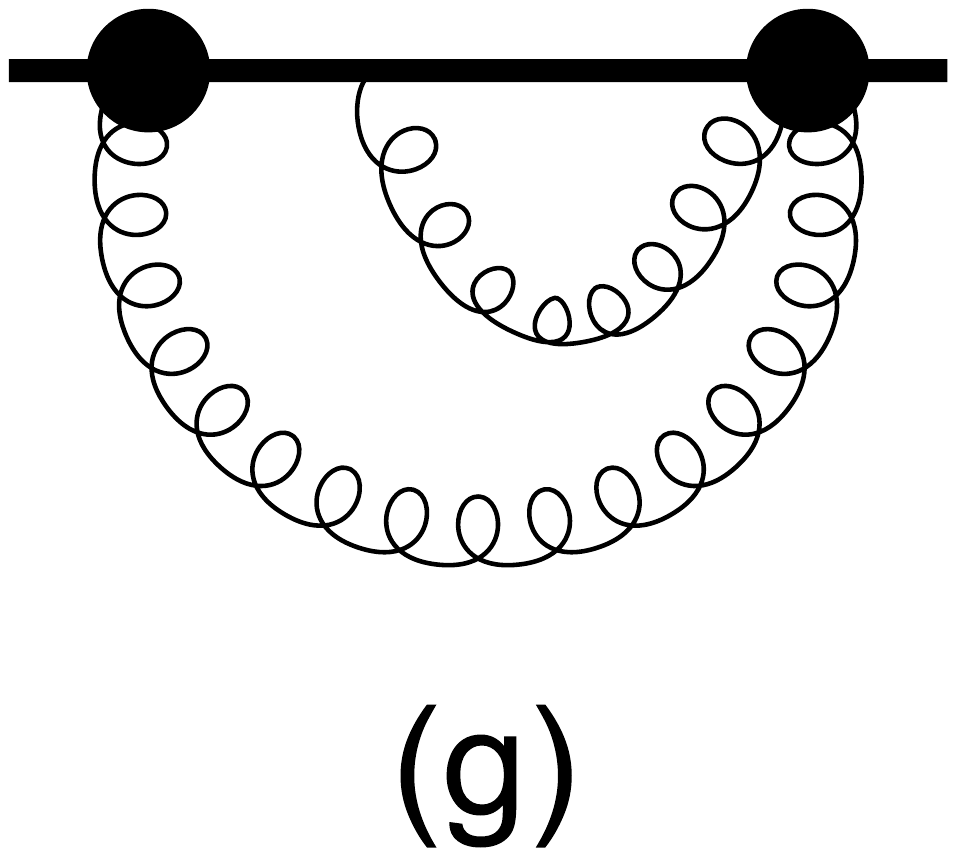} \hspace{-1.5em} & \includegraphics[scale=0.1]{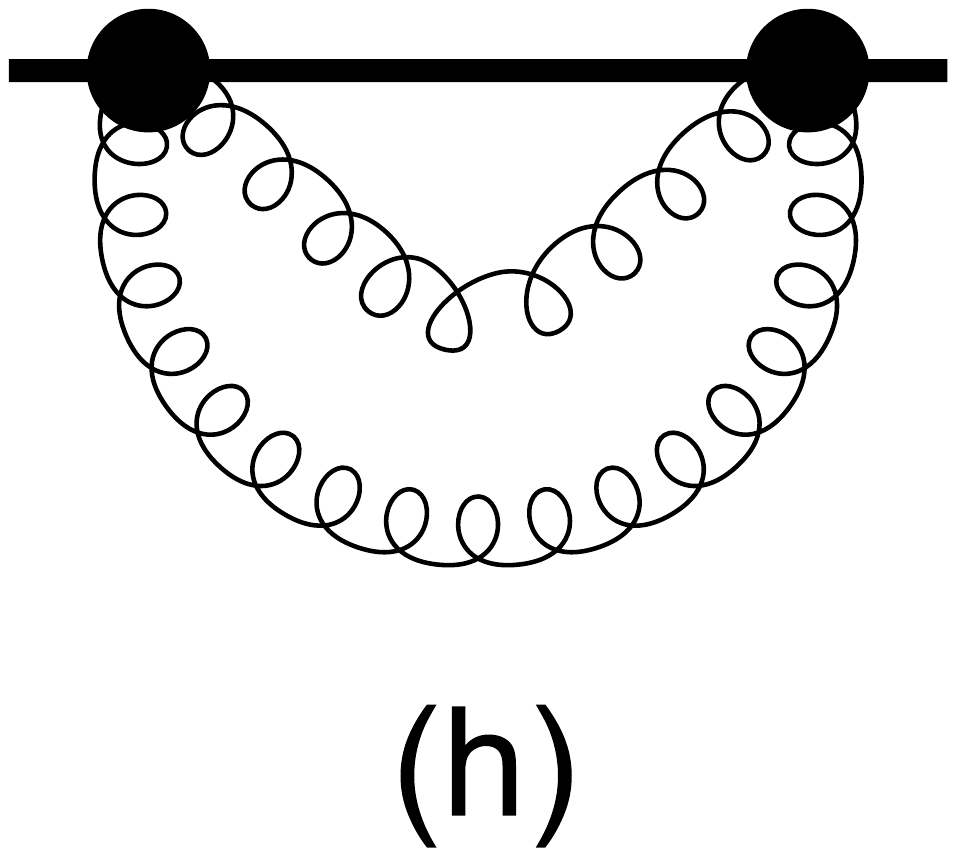}\vspace{-5ex} \tabularnewline  \includegraphics[scale=0.1]{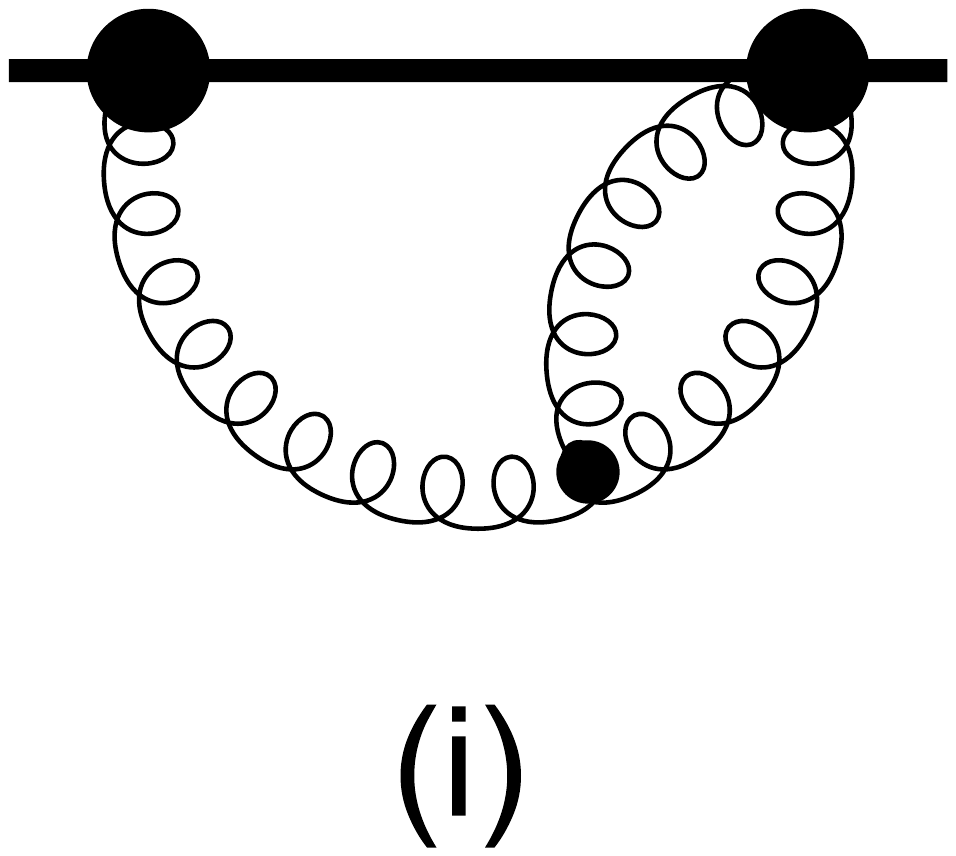} \hspace{-1em} &  \includegraphics[scale=0.1]{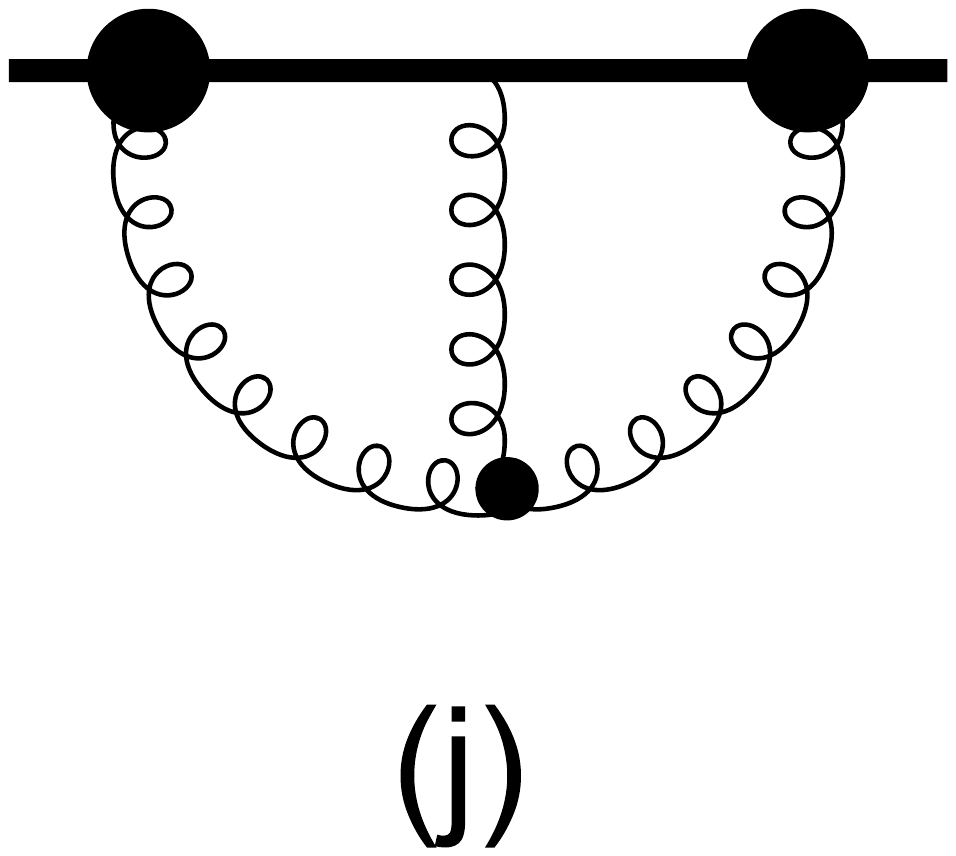} \hspace{-1em} &  \includegraphics[scale=0.1]{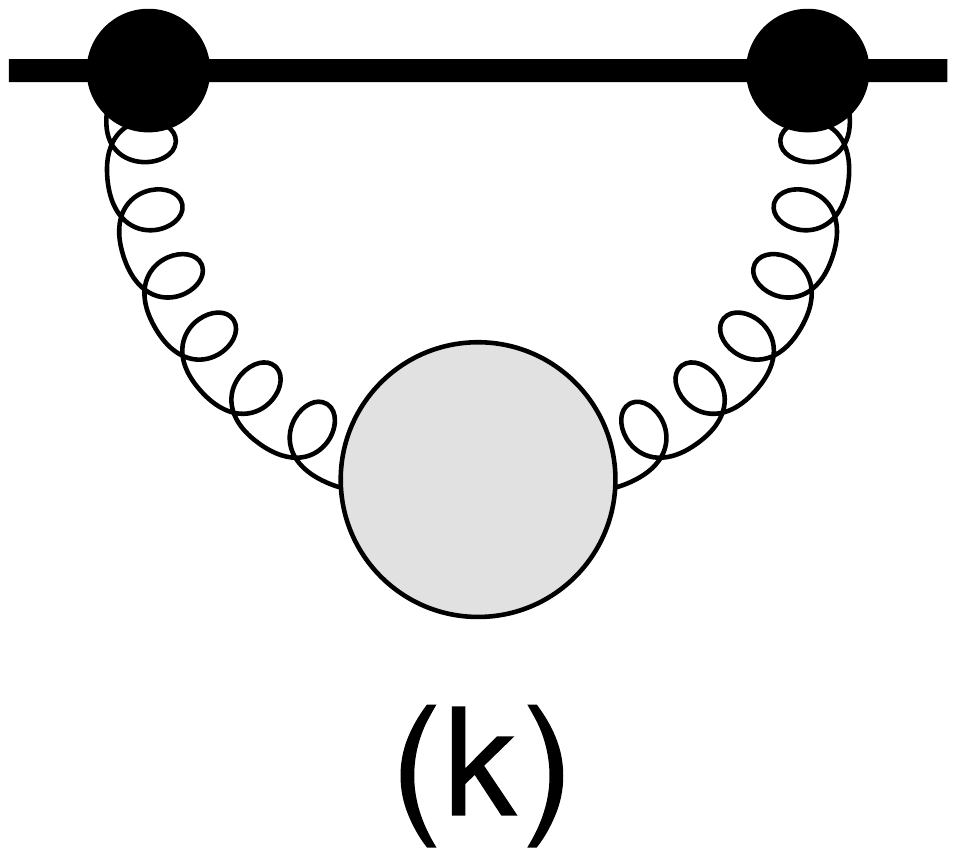} \vspace{-7ex}
\end{tabular}\caption{Diagrams without gradient flow in vacuum.
The naming convention follows Ref.~\cite{Burnier:2010rp}.
The field strength is represented by a big dot. Small dots are vertices and the grey blob the 1-loop self-energy. Curvy lines are gluon propagators and the Wilson line is a straight line.}
\label{Unflowed diagrams}
\end{figure}

The diagrams at NLO are presented in Figure \ref{Flowed diagrams}.
Similarly, Figure \ref{Unflowed diagrams} shows the diagrams in the unflowed $\MSBAR$ calculation.
Our naming convention follows Ref.~\cite{Burnier:2010rp},
but where the flow Feynman rules generate multiple diagrams with the same topology, we index them, eg, $(i_1),(i_2),(i_3)$ have the same topology as diagram $(i)$ of Ref.~\cite{Burnier:2010rp}.

To illustrate the calculation, we begin with diagram $(a)$:
\begin{align}
\GE^{\mathrm{{LO}}}\left(\tau,\tauf\right)
& = -\frac{\langle \mathrm{Re\: Tr}\left[g_{B} F_{0i}\left(x,\tauf\right) g_{B} F_{0i}\left(y,\tauf\right) \right]\rangle}{3\DR},
\nonumber \\ \label{diagramA}
\GB^{\mathrm{{LO}}}\left(\tau,\tauf\right)
& = \frac{\langle \mathrm{Re\: Tr}\left[g_{B} F_{jk}\left(x,\tauf\right) g_{B} F_{jk}\left(y,\tauf\right) \right]\rangle}{6\DR}.
\end{align}
We write the coupling as $g_B$ to emphasize that it is the bare coupling, which in an NLO calculation will contain the renormalized coupling and counterterms which we return to at the end of the calculation.
We also write for a general representation and group, with
$\CR$ the quadratic Casimir of the Wilson line representation,
$\DR$ the associated representation dimension, and
$\CA$ the quadratic Casimir of the adjoint representation.
For SU(3) and the fundamental representation they are
$\DR = 3$, $\CA=3$ and $\CR = 4/3$.
Writing out the field strengths at lowest order, this is
\begin{align}
\label{GEintermediate}
    \GE^{\mathrm{{LO}}}\left(\tau,\tauf\right)
    & =-\frac{g_{B}^{2}}{3\DR}\left\langle \mathrm{Re}\Tr\left[\left(\partial_{0}^{x}\delta_{i\alpha}-\partial_{i}^{x}\delta_{0\alpha}\right)B_{\alpha}^{m}\left(x,\tauf\right)\right.\right.
 \nonumber \\ &
    \left.\left.\times T^{m}\left(\partial_{0}^{y}\delta_{i\beta}-\partial_{i}^{y}\delta_{0\beta}\right)B_{\beta}^{n}\left(y,\tauf\right)T^{n}\right]\right\rangle
\\
\label{GBintermediate}
    \GB^{\mathrm{{LO}}}\left(\tau,\tauf\right) 
    & =\frac{g_{B}^{2}}{6\DR}\left\langle \mathrm{Re}\Tr\left[\left(\partial_{j}^{x}\delta_{k\alpha}-\partial_{k}^{x}\delta_{j\alpha}\right)B_{\alpha}^{m}\left(x,\tauf\right)\right.\right.
\nonumber \\ & 
    \left.\left.\times T^{m}\left(\partial_{j}^{y}\delta_{k\beta}-\partial_{k}^{y}\delta_{j\beta}\right)B_{\beta}^{n}\left(y,\tauf\right)T^{n}\right]\right\rangle \,. 
\end{align}
After taking the derivatives, we set $x=0$ and $y=(\tau,\vec 0)$.
The group trace is simple and gives a factor of $\CR \DR$.
Define the coordinate-space flowed-field propagator between two points $(x,y)$ and with flow depths $(s_1,s_2)$ as
\begin{equation}
    \left\langle B_{\mu}^{a}\left(x,s_1\right)B_{\nu}^{b}\left(y,s_2\right)\right\rangle
    =\delta^{ab}\delta_{\mu\nu}
    \Delta \left(x-y\right)_{s_1+s_2}.
\label{flowedcorrelator}
\end{equation}
Here $\Delta(x-y)_{s_1+s_2}$ is the scalar correlator in coordinate space after flow.
Then applying the derivatives shown in \Eq{GEintermediate} and \Eq{GBintermediate} yields
\begin{align}
\label{GEsecond}
 \GE^{\mathrm{{LO}}}\left(\tau,\tauf\right)
    & = -g_B^2 \CR \left( (D{-}1) \partial_0^x \partial_0^y + \partial_i^x \partial_i^y \right)
    \Delta\left(x{-}y \right)_{2\tauf},
    \\
    \label{GBsecond}
    \GB^{\mathrm{{LO}}}\left(\tau,\tauf\right)
    & = g_B^2 \CR \left( 2(D{-}2) \partial_i^x \partial_i^y \right)
    \Delta\left(x{-}y \right)_{2\tauf} \,.
\end{align}

We now need an explicit expression for the scalar propagator under gradient flow in coordinate space.
We find it by Fourier transforming the momentum-space expression:
\begin{align}
\label{scalarpropagator}
    & \Delta \left(x-y\right)_{s_1+s_2} 
    \\ & \equiv
    \int_{p}e^{ip\cdot(x-y)} \frac{e^{-(s_1+s_2)p^{2}}}{p^{2}}
\nonumber \\
    & = \frac{2^{D-2}}{(x-y)^{D-2} \left(4\pi\right)^{D/2}}
    \left(
    \Gamma \left(\textstyle{\frac D2}-1 \right)
    - \Gamma\left( \textstyle{\frac D2 -1 ,
    \frac{(x-y)^2}{4(s_1+s_2)}} \right) \right)
    \nonumber \\ \nonumber
    & \Rightarrow_{D=4}
    \frac{1}{4\pi^2 (x-y)^2} 
    \left( 1 - e^{-(x-y)^2/4(s_1+s_2)}  \right) .
\end{align}
Here $\Gamma(x)$ is the usual gamma function and $\Gamma(x,y)$ is the incomplete gamma function.

Note that in carrying out the integral in \Eq{scalarpropagator}, the relevant physical momentum scale $p$ will be the smaller of $1/|x-y|$ and $1/\sqrt{s_1+s_2}$.
The condition $p^2 \lesssim 1/(s_1+s_2)$ is clear from the exponential.
To see why $p$ must also be smaller than $\sim 1/|x-y|$, consider the case where $(x-y)$ is purely temporal.
Then one can push the $p_0$ integration contour into the positive imaginary half-plane.
The first singularity occurs at $\mathrm{Im}(p_0) = |\vec p|$,
arising from the $1/p^2$ factor.
If $|\vec p| |x-y| \gg 1$ then this gives rise to an exponential suppression of the $p_0$-integral.
Therefore the relevant $p$-integration range is $|p| \sim 1/|x-y|$.
This observation will be important later when we need to estimate the relevant ranges of different momenta in more complicated graphs.

To compute \Eq{GEsecond}, \Eq{GBsecond}, we need $s_1+s_2=2\tauf$, in which case the exponent above
is $\exp(-\chi)$.
Since we want $\chi \gg 1$, we can drop the incomplete gamma function, which makes the remaining steps very simple even without taking $D \to 4$.
One quickly finds
\begin{equation}
\label{GEfinal}
G_{E,B}^{\mathrm{{LO}}}\left(\tau,\tauf\right)=g_{B}^{2}\CR\frac{\left(D-1\right)\left(D-2\right)}{6}\frac{\tau^{-D}}{\pi^{\frac{D}{2}}}\Gamma\left(\frac{D}{2}\right),
\end{equation}
which, for $D=4$, gives $g_{B}^{2}\CR/\pi^{2}\tau^{4}$.

To extend the calculation to NLO, we need the NLO relationship between the bare and renormalized coupling: 
\begin{align}
\label{betafunc}
        g_{B}^{2} & \rightarrow g^{2}+\frac{g^{4}\beta_{0}}{\left(4\pi\right)^{2}}\left(-\frac{1}{\epsilon}+\mathcal{O}\left(\epsilon\right)\right),
 \\ \nonumber
        \beta_{0} & =\frac{11\CA -4N_{f} \TR}{3}.
\end{align}
Here $N_f$ is the number of Dirac fermionic flavors and $\TR$ is the trace normalization
$\TR = \CR \DR/D_{\mathcal{A}} = 1/2$ for the fundamental representation in QCD.
Inserting this expression into \Eq{GEfinal} requires the expansion of that expression to $\mathcal{O}\left(\epsilon=(4-D)/2 \right)$.
We do so within the $\MSBAR$ scheme and quote the result in Table \ref{TableFlowed}, along with all other NLO diagrams.

Let us quickly analyze the behavior of the remaining diagrams.
For a diagram to contain UV divergences, there must either be a loop at zero flow time as in $(k_{1\ldots 4})$ or there must be a loop in which the flow time is integrated over and it can extend down to zero,
as in $(k_{5\ldots 8})$.
Therefore only $k$-diagrams can contain UV divergences which must be regulated using $\MSBAR$.
All of these diagrams can be understood as self-energy corrections.
All other diagrams are finite in 4 dimensions, though in some cases the evaluation may be easier if we work in D dimensions during some intermediate steps.

Diagrams $(a)$ through $(h)$, which only involve lines originating from the Wilson line, turn out to be easiest to evaluate in coordinate space.
Since these diagrams are also UV finite, we can evaluate them in 4 dimensions by using the propagator expression in the last line of \Eq{scalarpropagator}.
For the correlator of two $B$ fields,
diagrams $(d,e,f,g)$ vanish because of the absence of a $\langle B_0 B_i \rangle$ correlator.
For two $E$ fields $(d,e,f)$ also vanish because they involve spatial dot products of independent loop momenta.

Diagrams $(b,c,g,j)$ depend on the location of attachment of the $A^0$ field or fields on the Wilson line due to group theory factors.
To illustrate that, first consider diagram $(g)$.
We must integrate over the attachment point, but the group-theory factor is opposite when it attaches between $(0,\tau)$ than when it attaches between $(-\infty,0)$ or $(\tau,\infty)$.
Therefore the diagram evaluates to
\begin{align}
\label{diagramg}
\nonumber & 
    \delta_{(g)} \GE^{\mathrm{NLO}}(\tau,\tauf) \\
    = & g^4\CR \CA
    \partial_\tau \frac{1}{4\pi^2 \tau^2}
    \times \nonumber \\ & 
    \int_{-\infty}^\infty d\ell
    \left( \Theta(-\ell) - \Theta(\ell) \Theta(\tau-\ell) + \Theta(\ell-\tau) \vphantom{\Big|} \right)
    \frac{1}{4\pi^2 \ell^2} 
    \nonumber \\
    = & g^4 \CR \CA
    \; \frac{-1}{2\pi^2 \tau^3} \;
    2\int_\tau^\infty d\ell \; \frac{1}{4\pi^2 \ell^2}
    \\ \nonumber
    = & - \frac{g^4 \CR \CA}{4\pi^4 \tau^4}.
\end{align}
Diagram $(h)$ is even easier, as it simply involves the square of the propagator between 0 and $\tau$.
The electric and magnetic versions have opposite signs due to the sign difference in the definitions of $\GE$ and $\GB$.

There is also a subtlety associated with diagrams $(b,c)$.
The difference between diagrams $(b)$ and $(c)$ is that in diagram $(c)$ one of the two attachments lies between the field strengths at $0$ and $\tau$, whereas for $(b)$ they either both lie between $0$ and $\tau$ or they both lie outside.
For diagram $(b)$ the group-theory factor is
$\Tr T^A T^A T^B T^B = \DR \CR^2$, while for diagram $(c)$ it is
$\Tr T^A T^B T^A T^B = \DR \CR^2 - \DR \CR \CA/2$.
The denominator in \Eq{Euclidean} or \Eq{BBEuclidean} introduces a graph of the identical form to $(b)+(c)$ but with a - sign (from expanding a denominator) and the same graph-theoretical factor as $(b)$.
Therefore the contribution from both diagrams, plus the counterterm, arises only from the attachment locations of diagram $(c)$:
\begin{align}
\label{diagramsbc}
    \delta_{(b,c)} G_{E,B}^{\mathrm{NLO}}
    (\tau,\tauf) = &
    (a) \times \frac{g^2 \CA}{2} \:
    2 \!\int_0^\tau \!\!\!d\ell_1 \!\!
    \int_\tau^\infty \!\!\! d\ell_2 \:
    \Delta(\ell_1{-}\ell_2).
\end{align}
If we use 
$\Delta(\ell_1-\ell_2)=1/4\pi^2 (\ell_1-\ell_2)^2$
in this expression then we leave in a logarithmic short-distance divergence, so we must use the full expression in the last line of \Eq{scalarpropagator}.
The integration is straightforward if we treat
$\tau^2/8\tauf \equiv \chi \gg 1$, leading to
\begin{equation}
\label{diagramsbc2}
  \delta_{(b,c)} G_{E,B}^{\mathrm{NLO}}
  (\tau,\tauf) = 
    (a) \times \frac{g^2 \CA}{8\pi^2}
    \left( \ln \frac{\tau^2}{8\tauf} + \gamma_E + 2 \right).
\end{equation}
The appearance of a logarithm in this diagram combination, as well as in $(i_1)$ and $(j_1)$, occurs because the diagrams are divergent in 4 dimensions in the absence of flow.
Flow naturally cuts off this divergence, leading to a logarithm $\ln(\tau^2/\tauf)$ with the same coefficient as the $1/\epsilon$ term in the unflowed $\MSBAR$ version of the diagram.

The remaining diagrams, namely the $(i,j,k)$ diagrams,
involve at least one vertex off the Wilson line and are best evaluated using momentum-space methods.
Their evaluation is considerably more involved than the above diagrams and we postpone it to an appendix.






\section{Diagram-by-diagram results}
\label{sec:results}

We saw in \Eq{GEfinal} that the leading-order electric and magnetic field strength correlators equal and are given by
\begin{equation}
    \GE^{\mathrm{LO}}(\tau,\tauf) = \GB^{\mathrm{LO}}(\tau,\tauf) = \frac{g^2 \CR}{\pi^2 \tau^4} \,.
\end{equation}
The typical NLO contribution is of form
\begin{align}
\delta_{(X)} G^{\mathrm{NLO}}
    & \simeq
    \frac{g^2 \CR}{\pi^2 \tau^4} \: \frac{g^2 \CA}{16\pi^2}
    \times 
    \left( c_1 \frac{1}{\epsilon'}
    + c_2 \left( \ln(\chi) + \gamma_E \right) + c_3 \right)
    \nonumber \\
    \frac{1}{\epsilon'} & \equiv
    \frac{1}{\epsilon} + 2 \ln \frac{\mubar^2 \tau^2}{4} + 4 \gamma_E \,.
\end{align}
with $c_1,c_2,c_3$ simple rational expressions.
We recognize the typical loop-suppression factor of $g^2 \CA/16\pi^2$, with the adjoint Casimir $\CA$ indicating purely nonabelian effects.
The coefficients $c_1,c_2,c_3$ are computed diagram by diagram and are tabulated in Table \ref{TableFlowed} for both the electric and the magnetic case.
There are two cases where the coefficients are more complicated.
The fermionic self-energy involves a different group Casimir and so does its associated counterterm, so we introduce
\begin{equation}
    F \equiv \frac{4 N_f \TR}{3\CA}
\end{equation}
to put it in the same form as the other terms.
And the NLO expression for diagram $(a)$ arising from the $\mathcal{O}(\epsilon)$ part in \Eq{GEfinal} multiplying the $1/\epsilon$ counterterm in \Eq{betafunc} gives rise to a $\mubar$ dependent and more complicated constant term
\begin{equation}
    a_1 \equiv \left( \frac{11}{3} - F \right)
    \left( \ln \frac{\tau^2 \mubar^2}{4} + 2 \gamma_E + \frac 83 \right).
\end{equation}
Also note that, for diagram $(j_1)$, the constant term contains a part proportional to $\pi^2$ as well as a simple rational number.

\begin{table*}[htb]
\centering{}%
\begin{tabular}{|c|c|c|c|c|c|c|}
\hline 
 & \multicolumn{1}{c}{} & \multicolumn{1}{c}{EE-Correlator} &  & \multicolumn{1}{c}{} & \multicolumn{1}{c}{BB-Correlator} & \tabularnewline
\hline 
Diagram & $\frac{1}{\epsilon'}$ & $\ln\left(\chi\right)+\gamma_{E}$ & Constant & $\frac{1}{\epsilon'}$ & $\ln\left(\chi\right)+\gamma_{E}$ & Constant\tabularnewline[0.1cm]
\hline 
$a$ & $F-\frac{11}{3}$ & 0 & $a_{1}$ & $F-\frac{11}{3}$ & 0 & $a_{1}$\tabularnewline[0.1cm]
\hline 
$b-f$ & 0 & $2$ & $4$ & 0 & $2$ & $4$\tabularnewline
\hline 
$g$ & 0 & 0 & $-4$ & 0 & 0 & 0\tabularnewline
\hline 
$h$ & 0 & 0 & $-1$ & 0 & 0 & $1$\tabularnewline
\hline 
$i_{1}$ & 0 & $-3$ & $3$ & 0 & $-3$ & 0\tabularnewline
\hline 
$i_{2}$ & 0 & 0 & $-\frac{25}{8}$ & 0 & 0 & $-\frac{25}{8}$\tabularnewline[0.1cm]
\hline 
$i_{3}$ & 0 & 0 & 0 & 0 & 0 & 0\tabularnewline
\hline 
$j_{1}$ & 0 & $3$ & $1-\frac{2\pi^{2}}{3}$ & 0 & $1$ & $5-\frac{2\pi^{2}}{3}$\tabularnewline[0.1cm]
\hline 
$j_{2}$ & 0 & 0 & $\frac{5}{2}$ & 0 & 0 & 0\tabularnewline[0.1cm]
\hline 
$j_{3}$ & 0 & 0 & $-\frac{3}{8}$ & 0 & 0 & $\frac{1}{8}$\tabularnewline[0.1cm]
\hline 
$k_{(1-4)}$ & $\frac{5}{3}-F$ & 0 & $-\frac{13}{3}+3F$ & $\frac{5}{3}-F$ & 0 & $-\frac{8}{3}+2F$\tabularnewline[0.1cm]
\hline 
$k_{5}$ & $-3$ & $3$ & $7$ & $-3$ & $3$ & $7$\tabularnewline
\hline 
$k_{6}$ & $2$ & $-2$ & $-\frac{35}{6}$ & $2$ & $-2$ & $-\frac{35}{6}$\tabularnewline[0.1cm]
\hline 
$k_{7}$ & $3$ & $-3$ & $\frac{-11}{2}$ & $3$ & $-3$ & $\frac{-11}{2}$\tabularnewline[0.1cm]
\hline 
$k_{8}$ & 0 & 0 & 0 & 0 & 0 & 0\tabularnewline
\hline 
\end{tabular}
\caption{Field strength correlator results for $\chi\rightarrow\infty$.
Each value is multiplied by  $g^{4}\CA\CR/\left(2\pi\tau\right)^{4}$.
The definitions of $1/\epsilon'$, $F$, and $a_1$ are given in the main text.
\label{TableFlowed}
}
\end{table*}

\begin{table*}[htb]
\centering{}%
\begin{tabular}{|c|c|c|c|c|c|c|}
\hline 
 & \multicolumn{1}{c}{} & \multicolumn{1}{c}{EE-Correlator} &  & \multicolumn{1}{c}{} & \multicolumn{1}{c}{BB-Correlator} & \tabularnewline
\hline 
Diagram & $\frac{1}{\epsilon}$ & $\ln(\frac{\bar{\mu}^{2}\tau^{2}}{4})+2\gamma_{E}$ & Constant & $\frac{1}{\epsilon}$ & $\ln(\frac{\bar{\mu}^{2}\tau^{2}}{4})+2\gamma_{E}$ & Constant\tabularnewline[0.15cm]
\hline 
$a$ & $F-\frac{11}{3}$ & $F-\frac{11}{3}$ & $\frac{8}{3}\left(\frac{11}{3}-F\right)$ & $F-\frac{5}{3}$ & $F-\frac{5}{3}$ & $\frac{8}{3}\left(\frac{5}{3}-F\right)$\tabularnewline[0.1cm]
\hline 
$b-f$ & $2$ & $4$ & $-\frac{4}{3}$ & $2$ & $4$ & $-\frac{4}{3}$\tabularnewline[0.1cm]
\hline 
$g$ & 0 & 0 & $-4$ & 0 & 0 & 0\tabularnewline
\hline 
$h$ & 0 & 0 & $-1$ & 0 & 0 & $1$\tabularnewline
\hline 
$i_{1}$ & $-3$ & $-6$ & $8$ & $-3$ & $-6$ & $5$\tabularnewline
\hline 
$j_{1}$ & $3$ & $6$ & $-4-\frac{2}{3}\pi^{2}$ & $1$ & $2$ & $\frac{10}{3}-\frac{2}{3}\pi^{2}$\tabularnewline[0.1cm]
\hline 
$k_{(1-4)}$ & $\frac{5}{3}-F$ & $\frac{10}{3}-2F$ & $-\frac{13}{3}+3F$ & $\frac{5}{3}-F$ & $\frac{10}{3}-2F$ & $-\frac{8}{3}+2F$\tabularnewline[0.1cm]
\hline  
\end{tabular}\caption{Field strength correlator results for $\tauf=0$. Each value is multiplied by  $g^{4}\CA\CR/\left(2\pi\tau\right)^{4}$.
}
\label{Table of comparison unflowed}
\end{table*}


Note that the factors of $1/\epsilon$ in the flowed case add up to zero, both in the electric and in the magnetic case.
This means that the flowed correlation function is finite after renormalization of the coupling has been taken into account.
This is expected, because gradient flow should provide properly renormalized operators.
In the electric case the coefficients of the $\ln(\chi)$ terms also add to zero, indicating that the result is independent of the choice of flow depth.
This is not the case for the magnetic correlator; at NLO the result \textsl{is} sensitive to the chosen flow depth, indicating that the magnetic field inserted on the Wilson line possesses an anomalous dimension.

The presence of an anomalous dimension means that the matching between the flowed and the $\MSBAR$ operator is nontrivial.
To determine it, and to make sure there is no similar effect for electric field operators, we also compute the unflowed, $\MSBAR$ values of all diagrams.
In this case, every diagram is of form
\begin{equation}
\delta_{(X)} G^{\mathrm{NLO}}
\simeq \frac{g^4 \CR \CA}{16\pi^4 \tau^4} \times
    \left( c_1 \frac{1}{\epsilon} 
    + c_2 \left( \ln\frac{\mubar^2 \tau^2}{4} + 2 \gamma_E \right) + c_3 \right)
\end{equation}
with some coefficients $(c_1,c_2,c_3)$.
These coefficients are listed in Table \ref{Table of comparison unflowed}.
In evaluating the NLO contribution from diagram $(a)$, we must include both the gauge-coupling counterterm and a counterterm $2/\epsilon$ from the operator anomalous dimension, as otherwise the $1/\epsilon$ terms will not all cancel out.
The presence of an anomalous dimension, and its value, was already found by Bouttefeux and Laine \cite{Bouttefeux:2020ycy}, and our result agrees with theirs.

Previously, Laine and collaborators \cite{Burnier:2010rp,Laine:2021uzs}
have computed the same correlation functions, including finite-temperature effects; we examine the vacuum values from their work.
They evaluate the spectral function in the frequency domain, rather than the Euclidean correlator in the time domain.
The relation (in vacuum) is
\begin{equation}
    G(\tau) = \int_0^\infty \frac{d\omega}{\pi} \rho(\omega) e^{-\omega \tau}.
\end{equation}
If $\rho(\omega)$ is
$\rho(\omega) = \omega^3(A+B \ln(\omega^2/\mu^2))$,
one then finds
\begin{align}
    G(\tau) & = \int_0^\infty \frac{d\omega}{\pi}
    e^{-\omega \tau} \omega^3 \left( A + B \ln(\omega^2/\mu^2) \right)
    \nonumber \\ & =
    \frac{1}{\pi \tau^4} \int_0^\infty dy e^{-y} y^3
    \left(A + B \ln(y^2) -B \ln(\tau^2 \mu^2) \right)
    \nonumber \\ & =
    \frac{\Gamma(4)}{\pi \tau^4} \left( A - B \ln(\tau^2 \mu^2) + 2 B \psi(4) \right)
    \\ & = \frac{6}{\pi \tau^4} \left( A +
    B \left[ \frac{11}{3} - 2 \gamma_E - \ln(\tau^2 \mu^2) \right] \right),
\end{align}
where $\psi(x) = d(\ln\Gamma(x))/dx$ is the digamma function.
Using this relation, we find that our results agree with the NLO vacuum spectral functions from both references except for the $\pi^2$ term, where our coefficient $-2\pi^2/3$ appears as $-8\pi^2/3$ in both references.
This appears to arise from a difference in the analysis of diagram $(j)$ in the unflowed case and $(j_1)$ in the flowed case.
Recently Ref.~\cite{Scheihing_Hitschfeld_2023} found precisely this same factor-of-4 discrepancy with respect to Ref.~\cite{Burnier:2010rp}.
We explain the evaluation of this diagram in detail in Appendix \ref{secDetails}.

\section{Combining the results}
\label{sec:combining}

The explicit expressions of each scheme at NLO are summarized as follows:
\begin{widetext}
\begin{align}
\label{GEflow}
\GE^{\mathrm{{flow}},\bar{\mu}}\left(\tau\right) & 
= \frac{g^2 \CR}{\pi^2 \tau^4} \left[ 1 + \frac{g^2 \CA}{16\pi^2} \left[
\left(\frac{11}{3}-F\right)\left(\ln\left(\frac{\tau^{2}\bar{\mu}^{2}}{4}\right)+2\gamma_{E}-\frac{1}{3}\right)+\frac{13}{3}-\frac{2\pi^{2}}{3}
\right] \right] \\
\label{GBflow}
\GB^{\mathrm{{flow}},\bar{\mu}}\left(\tau\right) & 
= \frac{g^2 \CR}{\pi^2 \tau^4} \left[ 1 + \frac{g^2 \CA}{16\pi^2} \left[
\left(\frac{11}{3}-F\right)\left(\ln\left(\frac{\tau^{2}\bar{\mu}^{2}}{4}\right)+2\gamma_{E}+\frac{2}{3}\right)-2\left(\ln\left(\frac{\tau^{2}}{8\tauf}\right)+\gamma_{E}\right)+\frac{22}{3}-\frac{2\pi^{2}}{3}
\right] \right] \\
\label{GEmsbar}
\GE^{\overline{\mathrm{{MS}}},\bar{\mu}}\left(\tau\right) &
= \frac{g^2 \CR}{\pi^2 \tau^4} \left[ 1 + \frac{g^2 \CA}{16\pi^2} \left[
\left(\frac{11}{3}-F\right)\left(\ln\left(\frac{\tau^{2}\bar{\mu}^{2}}{4}\right)+2\gamma_{E}-\frac{1}{3}\right)+\frac{13}{3}-\frac{2\pi^{2}}{3}
\right] \right] \\
\label{GBmsbar}
\GB^{\overline{\mathrm{{MS}}},\bar{\mu}}\left(\tau\right) & 
= \frac{g^2 \CR}{\pi^2 \tau^4} \left[ 1 + \frac{g^2 \CA}{16\pi^2} \left[
\left(\frac{11}{3}-F\right)\left(\ln\left(\frac{\tau^{2}\bar{\mu}^{2}}{4}\right)+2\gamma_{E}+\frac{2}{3}\right)-2\left(\ln\left(\frac{\tau^{2}\bar{\mu}^{2}}{4}\right)+2\gamma_{E}\right)+\frac{22}{3}-\frac{2\pi^{2}}{3} \right] \right] .
\end{align}
\end{widetext}
To turn these expressions into expressions for the spectral function, replace $1/\pi^2 \tau^4 \to \omega^3/6\pi$
and $\ln(\tau^2 \mubar^2/4)+2\gamma_E \to \ln(\mubar^2/4\omega^2) + 11/3$.

The term starting with $(11/3 - F)$ represents the coupling renormalization, effectively fixing the coupling at the scale of the operator separation $\mubar \sim 1/\tau$.
We see that $\GE$ is the same in each scheme, does not depend on $\tauf$, and only depends on the renormalization point $\mubar$ through coupling renormalization.
The electric field inserted on a Wilson line does not require renormalization and its normalization is not affected by gradient flow.

In contrast, the magnetic correlators differ; we have
\begin{equation}
\label{premainresult}
    \frac{\GB^{\mathrm{flow}}}{\GB^{\MSBAR}} = 1 + \frac{g^2 \CA}{16\pi^2}
\: 2 \, \left( \ln (2\tauf \mubar^2) + \gamma_E  \right)
+ \mathcal{O}\left(g^4 \right)\,.
\end{equation}
This arises because the magnetic field insertion renormalizes and its normalization is scheme and scale dependent.
That is, we can define a relation between the magnetic field inserted on a Wilson line and renormalized via flow and the same structure renormalized using the $\MSBAR$ scheme as:
\begin{align}
\label{mainresult}
    \nonumber
    F_{ij}^{\mathrm{flow}}(\tauf) & = Z(\tauf,\mubar) F_{ij}^{\MSBAR}(\mubar) \,,
    \\
    Z(\tauf,\mubar) & = 1 + \frac{g^2 \CA}{16\pi^2}
    \left( \ln \frac{8\tauf \mubar^2}{4} + \gamma_E \right)
    +\mathcal{O}\left(g^4 \right)\,.
\end{align}
This relation, or equivalently \Eq{premainresult}, is the main result of this paper.
As mentioned before, it was already quoted by us, without computational details, in Refs.~\cite{Altenkort:2023eav,Altenkort:2024spl},
and has recently been derived independently in Ref.~\cite{Brambilla:2023vwm}.

In writing \Eq{mainresult}, we have made an arbitrary decision to write $F^{\mathrm{flow}}$ in terms of $F^{\MSBAR}$ and not \textsl{vice versa}.
Writing $Z(\tauf,\mubar) = 1+\kappa$ in \Eq{mainresult},
we could equally well have written
$F_{ij}^{\MSBAR} = (1-\kappa) F_{ij}^\tauf$.
Or we could have written $Z(\tauf,\mubar)=(1+2\kappa)^{1/2}$,
or $Z(\tauf,\mubar) = \exp(\kappa)$,
or any of a number of other choices.
These all differ from each other at the NNLO level, and without an NNLO calculation it is not clear to us which is preferred.

\Eq{mainresult} also shows that $F^{\MSBAR}_{ij}$ is scale dependent.
$F^{\mathrm{flow}}_{ij}$ is $\mubar$ independent, as one sees from the $\mubar$ independence of \Eq{GBflow} (the explicit $\mubar$ dependence cancels the $\mubar$ dependence of the gauge-coupling).
Therefore the product $Z(\tauf,\mubar) F^{\MSBAR}_{ij}$ is $\mu$ independent, and therefore
\begin{equation}
\label{CallanSymanzik}
    \frac{\mubar d}{d\mubar} F_{ij}^{\MSBAR}(\mubar)
    = - \frac{g^2 \CA}{8\pi^2} F_{ij}^{\MSBAR}(\mubar),
\end{equation}
which is the (already-known \cite{Bouttefeux:2020ycy}) Callan-Symanzik equation for the evolution of the field strength operator.
This can be used to renormalization-group evolve $F_{ij}^{\MSBAR}$ between different $\mubar$ values.

Also, because $F_{ij}^{\MSBAR}$ is scale dependent, there must be a $\mubar$ dependent relation between its correlator and the actual velocity-dependent force felt by a heavy quark.
Call the correctly-normalized field-strength correlator to use in determining $\kappa_B$ of \Eq{fullkappa} $\GBphysical$.
Laine \cite{Laine:2021uzs} has determined that, at NLO,
\begin{equation}
\label{GBphysical}
\GBphysical \left(\tau\right)
=\left(1+\frac{g^{2}\CA}{8\pi^{2}}\left[\ln\frac{\mubar^2}{(4\pi T)^2}+2\gamma_E -2\right] \right) \GB^{\MSBAR}(\mubar).
\end{equation}
This should be combined with \Eq{premainresult} to relate the flowed correlator, determined on the lattice, with the physical correlator needed for momentum diffusion.

To be more precise, we advocate choosing two renormalization scales:
$\muIR$, close to the scale $2\pi T$ where the matching in \Eq{GBphysical} naturally occurs, and $\muUV$, close to the scale
$1/\sqrt{\tauf}$ where the matching in \Eq{premainresult} naturally occurs.
One then applies \Eq{premainresult} at $\muUV$ to convert lattice results into $\MSBAR$ renormalized results;
renormalization-group flows between the scales $\muUV$ and $\muIR$ using \Eq{CallanSymanzik}; 
and matches to the physical value using \Eq{GBphysical}.
Minimizing NNLO effects by performing both matchings using an exponential version of \Eq{GBphysical} and \Eq{premainresult}, one obtains:
\begin{align}
\label{allZ}
    \GBphysical = {} & Z_{\mathrm{phys}}(\muIR,\muUV,\tauf)
    \GB^{\mathrm{flow}}(\tauf) \,, \\
    \ln
    Z_{\mathrm{phys}}( \ldots )
    = {} & 
    \frac{g^2(\muIR) \CA}{8\pi^2} \left[ \ln \frac{\muIR^2}{(4\pi T)^2} + 2 \gamma_E - 2 \right]
    \nonumber \\ &
    \left.
    + \int_{\muIR^2}^{\muUV^2} \frac{g^2(\mubar)\CA}{8\pi^2}
    \frac{d\mubar^2}{\mubar^2}
    \right. \nonumber \\ &
    - \frac{g^2(\muUV) \CA}{8\pi^2} \left[ \ln \frac{8\tauf \muUV^2}{4} + \gamma_E \right]
    . \nonumber
\end{align}
(As a check, the explicit $\muIR$ and $\muUV$ dependences from the two logs and from the limits of the integrals cancel, leaving only formally higher-order dependence from coupling renormalization.)
This translation should be applied \textsl{before} extrapolating to zero flow depth $\tauf \to 0$, that is, one should extrapolate $\GBphysical$ and not $\GB^{\mathrm{flow}}$ to $\tauf \to 0$, because $\GBphysical$ has no \textsl{expected, logarithmic} dependence on flow depth, only polynomially-suppressed $\tauf$ dependence arising from high-dimension operator contamination which should be extrapolated away.

The proposal to use \Eq{allZ} as the main matching tool between (lattice-evaluated) gradient-flowed magnetic field correlators and the physical correlation function to use in analytical continuation for the spectral function is the main \textsl{practical proposal} arising from this work.
It has already been implemented in Refs \cite{Altenkort:2023eav,Altenkort:2024spl}, which quoted the above results ahead of time.

\section*{Acknowledgements}
The authors acknowledge the support from the Deutsche Forschungsgemeinschaft (DFG, German Research Foundation) through the CRC-TR 211 'Strong-interaction matter under extreme conditions'– project number 315477589 – TRR 211.
We thank Panayiotis Panayiotou for an enlightening discussion.

\appendix

\section{Feynman rules of QCD and gradient flowed QCD in Euclidean spacetime}
\label{secFeynRules}

\input{appendixFeynRules}

\section{Details for (i),(j),(k) diagrams}
\label{secDetails}

\input{appendixDetails}



\medskip

\section{Euclidean correlator for quarkonium transport}
\label{secadjointline}

This paper explored electric fields along a fundamental-representation Polyakov loop because of its importance for open heavy flavor.
Quarkonium can exist in the singlet or adjoint representations, and the transitions between them may be related to the correlation function of two electric fields with an adjoint line running between them.
This is explored, for instance, in Refs.~\cite{Binder:2021otw,Scheihing-Hitschfeld:2022xqx,Scheihing_Hitschfeld_2023}.
The formal definition of the correlator is
\begin{align}
G_{E_{\mathrm{{adj}}}}\left(\tau\right)
& \equiv \frac{g^{2}T_{F}}{3N_{c}}\left\langle E_{i}^{a}\left(\tau\right)W^{ab}\left(\tau,0\right)E_{i}^{b}\left(0\right)\right\rangle
\\ \nonumber
W^{ab}(\tau,0) & =
\left[ \mathcal{P}\exp \left( ig\int_{0}^{\tau} ds \: A_{0}^{c} \left(s\right) T_{\mathrm{adj}}^{c}\right)\right]^{ab}.
\end{align}
This can differ from \Eq{Euclidean} due to diagrams with a line connecting to the Wilson line, that is, diagrams $(b,c,g,j)$.
(Diagrams $(d,e,f)$ still vanish.)
For the above correlator, gauge fields only connect to the Wilson line between $0$ and $\tau$,
rather than connecting in this region and connecting outside this region but with opposite sign.
That is, for diagrams $(g,j)$, we replace
$\Theta(-s)-\Theta(s)\Theta(\tau-s)+\Theta(s-\tau)$ in \Eq{j1int} with $-2\Theta(s) \Theta(\tau-s)$.
The difference between these expressions, after $s$-integration, is a factor of $-2\pi \delta(k^0)$.
Furthermore, the subtraction term which effectively removes $(b)$ and the fundamental-Casimir part of $(c)$ is not subtracted -- indeed, it's not clear how to evaluate it.
Instead, diagrams $(b,c)$ are replaced with only the part with connections between the field-strength insertions, and no subtraction.
The result is that there are now contributions which scale as $\tauf^{-1/2}$ and grow linearly with $\tau$.
In the hopes that it is helpful for subtracting these terms and understanding the complicated $\tauf$ dependence of this correlator, we calculate these diagrams here.

The change in Diagrams $(b),(c)$ is
\begin{equation}
    G_{\!\mathrm{E,B:adj}} - G_{\!\mathrm{E,B:fund}}
    = (a) \times \frac{\CA g^2}{16\pi^2} \times
    (-2) \sqrt{\frac{2\pi \tau^2}{\tauf}}.
    \label{divergentpart}
\end{equation}
This is the main new contribution, and we believe that it exponentiates when higher-order Wilson-line corrections are included, that is,
$G_{\mathrm{adj}} = G_{\mathrm{fund}} \times 
\exp(-2 \CA g^2 \sqrt{2\pi \tau^2/\tauf}/16\pi^2)$.
This is a large, linear-in-$\tau$ (perimeter-law), $\tauf$-dependent rescaling of the adjoint correlator,
representing a $\tauf$-dependent mass correction for the heavy-quark bound state
which must be taken into account when trying to extrapolate to the small $\tauf$ limit.
The presence and coefficient of this term is the main result of this appendix.

In addition, we find that the constant proportional to $\pi^2$ in diagram $(j_1)$ changes, so that $-2\pi^2/3$ appearing in Table \ref{TableFlowed} becomes $4\pi^2/3$.
This shift agrees with a shift found by Scheihing-Hitschfield and collaborators~\cite{Scheihing_Hitschfeld_2023}.

In addition, for the $E$-field case, we find corrections in diagrams $(g)$, $(j_1)$, $(j_2)$, and $(j_3)$ which are of the same form as \Eq{divergentpart}, but with the coefficient $(-2)$ replaced by $+1$, $-4/3$, $+2/3$, and $-1/3$ respectively.
The sum of these terms gives zero, and similar terms are absent for the $B$-field case.
We have not checked whether there is a gauge choice where these terms do not arise.


\bibliography{refs}

\end{document}

%% file: appendixFeynRules.tex
We will only present the Feynman rules in Feynman gauge, both for the flowed and the unflowed fields.

The flowed theory contains four kinds of gauge-field propagators.
The correlator of two unflowed fields is, as usual,
\begin{equation}
\left\langle A_{\mu}^{a}A_{\nu}^{b}\right\rangle (p)  = \delta^{ab}
\frac{\delta_{\mu\nu}}{p^2} \,,
\end{equation}
where we have used Feynman gauge, as we will throughout.
If one gauge field is at flow depth $\tau_1$ and the other is unflowed, the correlator is
\begin{equation}
\left\langle B_{\mu}(\tau_1) A_\mu \right\rangle (p) = \delta^{ab}
\frac{\delta_{\mu\nu} e^{-\tau_1 p^2}}{p^2} \,,
\end{equation}
and if both are flowed we have
\begin{equation}
\left\langle B_{\mu}(\tau_1) B_\mu(\tau_2) \right\rangle (p) = \delta^{ab}
\frac{\delta_{\mu\nu} e^{-(\tau_1+\tau_2) p^2}}{p^2} \,.
\end{equation}
Finally, the correlator between $B_\mu(\tau_1)$ and $L_\nu(\tau_2)$ the Lagrange-multiplier field,
where $\tau_2 < \tau_1$ is required, is
\begin{equation}
  \left\langle B_{\mu}(\tau_1) L_\mu(\tau_2) \right\rangle (p) = \delta^{ab}
  \delta_{\mu\nu} e^{-(\tau_1-\tau_2) p^2} \,.
\end{equation}
Note that the flow difference, not sum, appears, and there is no denominator.

Ghosts and fermions only exist at zero flow-time and they satisfy the usual propagator and vertex Feynman rules, which we will not list here.

The unflowed vertex Feynman rule is as usual
\begin{align}
\nonumber & V^{abc}_{\mu\nu\rho}(p,k,q)
\\ = & ig f^{abc}\left[\left(k-q\right)_{\mu}\delta_{\nu\rho}+\left(q-p\right)_{\nu}\delta_{\mu\rho}+\left(p-k\right)_{\rho}\delta_{\mu\nu}\right].
\end{align}

Four-gluon vertex action is
\begin{align}
W_{\mu\nu\rho\gamma}^{abcd}\equiv g^{2} \Big[ & f^{ebc}f^{ead}\left(\delta_{\mu\nu}\delta_{\rho\gamma}
-\delta_{\mu\rho}\delta_{\nu\gamma}\right) 
\nonumber \\ &
+ f^{ebd}f^{eac}\left(\delta_{\mu\nu}\delta_{\rho\gamma}
- \delta_{\mu\gamma}\delta_{\nu\rho}\right)
\nonumber \\ &
+f^{eba}f^{edc}\left(\delta_{\mu\rho}\delta_{\nu\gamma}
-\delta_{\mu\gamma}\delta_{\nu\rho}\right) \Big].
\label{Four-gluon vertex action}
\end{align}

The flowed three-gluon vertex arises at a flow depth $s$ which must be integrated over.
It always involves one $L$ field, which must correlate with a $B$ at a larger flow time, and two $B$ fields.
In Feynman gauge it differs from the vacuum form:
the vertex between $L_\mu^a$ and $B_\nu^b$, $B_\rho^c$ is
\begin{align}
igf^{abc}
\left[\left(q-p-k\right)_{\nu}\delta_{\mu\rho}+\left(p-k+q\right)_{\rho}\delta_{\mu\nu}+\left(k-q\right)_{\mu}\delta_{\nu\rho}\right].
\end{align}
The flowed four-gluon vertex also contains one $L$ and three $B$ fields, but its form is the same as the vacuum one.

%% file: appendixDetails.tex
In the main text we showed how diagrams $(a)$ through $(h)$ can be carried out easily using real-space methods.
Diagrams $(i),(j),(k)$ are more involved, so we present the details in this appendix.
It is not our intention to provide \textsl{every} detail, but we will show how to handle the most sensitive steps, and we will give considerable detail to the parts of diagram $(j_1)$ where our results differ from Burnier et al.

We start with $(i)$, with one 3-gluon vertex and one two-field field strength.
Under gradient flow this becomes three diagrams;
diagram $(i_1)$, with an unflowed vertex;
diagram $(i_2)$, where the vertex occurs in flowing one of the two $B$-fields from the two-field field-strength,
and diagram $(i_3)$, where the vertex arises in flowing the $\partial_i B_j$ type field strength.
As mentioned in the main text, every loop integral has an exponential suppression factor with a finite flow depth which cannot approach zero.
Therefore each diagram is finite.
Nonetheless we will work in $D$ dimensions during intermediate steps, in case manipulations split the result into divergent terms with cancelling divergences.

\begin{widetext}

The Feynman rules of the previous appendix applied to diagram $(i_1)$ produce a contribution of
\begin{align}
\label{i1int}
\delta_{(i_1)} G(\tau,\tauf) & =
g^{4} \CA\CR \:
\int_{p,k}e^{ip_0 \tau}
\frac{e^{-\tauf(p^{2}+(p-k)^{2}+k^{2})}}{p^{2}\left(p-k\right)^{2}k^{2}}
\times \left\{
\begin{array}{ll}
(D-1) p_0^2 + p_j^2 & E \\
-(D-2) p_j^2 & B \\ \end{array}
\right\}.
\end{align}
To perform the $k$ integration, we exponentiate the denominators using
\begin{equation}
\label{howtoexponentiate}
    \frac{e^{-\tauf k^2}}{k^2} = \int_\tauf^\infty e^{-x k^2} dx.
\end{equation}
All $k$-factors then appear in the exponential, where one can shift the integration variable and perform the simple Gaussian $k$ integration:
\begin{align}
\label{dothek}
\int_k \frac{e^{-\tauf(k^2 + (k-p)^2)}}{k^2(k-p)^2}
& = \int_k \int_\tauf^\infty dx dy \;
e^{-(x+y)k^2} e^{-\frac{xy}{x+y} p^2}
\\
& = \frac{1}{16\pi^2} \int_1^\infty dx' dy' \;
\frac{e^{-\frac{x'y'}{x'+y'} \tauf p^2}}{(x'+y')^2}
\nonumber \\
& = \frac{1}{16\pi^2} \left[
 \mathrm{Ei}\left( \frac{-\tauf p^2}{2} \right)
 - 2 \mathrm{Ei}(-\tauf p^2) 
 + \frac{2}{\tauf p^2} \left( e^{-\tauf p^2/2} - e^{-\tauf p^2} \right) \right].
\end{align}
Here $\mathrm{Ei}(x)$ is the exponential integral function.
The resulting $p$ integral is also solvable in closed form for general $\tauf/\tau^2$.
We only need the (rather simpler) large $\tau^2/\tauf$ limit, which is given in Table \ref{TableFlowed}.
The result is logarithmically large.
We can already see the origin of this log in \Eq{dothek}; the $k$-integration is of form $d^4 k / k^4$, cut off in the UV at the scale $k^2\sim 1/\tauf$ by the exponential factor and cut off in the infrared by the appearance of $p$ in the denominator.
As mentioned in the main text, the $e^{ip_0\tau}/p^2$ factor ensures that $|p| \sim 1/\tau \ll 1/\sqrt{\tauf}$.
Hence the IR and UV cutoffs on the $k$-integration are at the well-separated scales $1/\tauf$ and $1/\tau^2$ respectively, leading to the logarithm in $\chi = \tau^2/8\tauf$.

Diagram $(i_2)$ does not produce such a logarithm, because one of the propagators in the $k$-loop is a $BL$-type propagator without a denominator.  In 4 dimensions it gives
\begin{equation}
    \label{i2int}
\delta_{(i_2)} G(\tau,\tauf) = 
\frac{2g^{4} \CA\CR}{3}\mbox{Re}\int_{0}^\tauf ds
\int_{p,k}e^{ip_{0}\tau} \frac{e^{-(\tauf+s)(p^{2}+k^{2})}}{p^{2}k^{2}} e^{-(\tauf-s)(p+k)^{2}}
\left\{
\begin{array}{ll}
3p_0(3p_0-k_0) + p_i (3p_i-k_i) & E \\
-p_i (3p_i-k_i) & B \\
\end{array}
\right\}.
\end{equation}
We see that the $k$-integration is of form $d^4 k / k^2$ and it is therefore dominated by the UV scale $k^2 \sim 1/\tauf$.
There is therefore a hierarchy, $p \ll k$, and we can expand
$e^{-(\tauf - s)(p+k)^2} \simeq e^{-(\tauf-s) k^2}
(1 -2(\tauf-s) p_\mu k_\mu)$.
The $p_\mu k_\mu$ term gives nonzero results when multiplying the linear-in-$k$ terms in the last expression, while the $1$ term gives nonzero results when multiplying the $k$-independent terms in the last expression.
Using $\int_k k_\mu k_\nu = \frac{\delta_{\mu\nu}}{4} \int_k k^2$, we can perform the $k$-integral, leaving $p$-integrals which can be performed with the same methods as for diagram $(a)$.

An integral expression for the $(i_3)$ diagram is:
\begin{equation}
\delta_{(i_{3})}G\left(\tau,\tauf\right) = \frac{2 g^{4} \CA\CR}{3} \mbox{Re}\int_{0}^\tauf ds\int_{p,k}e^{ip_{0}\tau}\frac{e^{-(\tauf+s)(k^{2}+(p+k)^{2})}}{k^{2}\left(p+k\right)^{2}}e^{-(\tauf-s)p^{2}}
\left\{ \begin{array}{ll} 
3 p_0^2 + p_i^2 & E \\
2 p_i^2 & B \\
\end{array} \right\}.
\label{i3int}
\end{equation}
This expression is strictly subleading in the large-$\chi$ expansion, so for our purposes it evaluates to 0.
To see this, note that the $p_0$ integral can again be pushed into the upper imaginary plane until it encounters the first singularity.
But there is no $1/p^2$ term; singularities only arise from $1/k^2 (p+k)^2$.
If $k$ is large, the first singularity occurs for $p \sim k$, leading to an exponential $e^{-k\tau}$ suppression; therefore \textsl{both} the $(p,k)$ integration variables must be $\sim 1/\tau$.
But the $\int_0^\tauf ds$ integral gives rise to an explicit $\sim \tauf$ term, leading to a $\tauf/\tau^2$ suppression.
Therefore the result is smaller than $(i_1)$ by a factor of order $\tauf/\tau^2 \sim 1/\chi$.

A more physical way to see this is to think about position space.
The field strength with a single $B$-field attachment is connected to the 3-point vertex by a purely flow-time propagator.
The flow smearing radius is of order $r \sim \sqrt{\tauf}$, so the 3-point vertex is within $\sqrt{\tauf}$ of the field-insertion in real space (to avoid an exponential penalty).
Therefore the two propagators in the $k$-loop must propagate a distance of order $1/\tau$.
This constrains \textsl{both} $k$ and $p$ to be of order $1/\tau$.
But the flow-time integration $\int_0^\tauf  ds$ introduces a positive power of $\tauf$, which cannot be compensated since all other objects are at the $\tau$ scale.

Applying the Feynman rules to the $(j_{1})$ diagram, and defining $p$ to be the momentum arriving at the field strength tensor at $\tau$ and $k$ to be the momentum arriving at the vertex on the Wilson line, we obtain
\begin{align}
\label{j1int}
    \delta_{(j_1)} G(\tau,\tauf) = &
    \frac{ig^4 \CA \CR}{6} \int_{pk} \frac{e^{ip_0\tau} e^{-\tauf (p^2+k^2+(p+k)^2)}}{p^2 k^2 (p+k)^2}
    \int_{-\infty}^{\infty} ds e^{ik_0 s}
    \left( \Theta(-s) - \Theta(s) \Theta(\tau-s) + \Theta(s-\tau) \right)
    \nonumber \\ & \hspace{4em}
    \times \left\{ \begin{array}{ll}
     -(D-1)p_0(p_0+k_0)(2p_0+k_0)
     - 2p_0(p_i^2 + p_i k_i + k_i^2) - k_0 (p_i^2 - p_i k_i)
     & E \\
     (D-2) (2p_0+k_0) p_i (p_i+k_i) & B \\
    \end{array} \right\} \,.
\end{align}
The integral over $s$ represents the location along the Wilson line where the gauge field attaches.
The combination of $\Theta$ functions arises because the group-theory factor is the opposite when the Wilson-line attachment is between the two field-strengths as when it is before or after both.

Like diagram $(i_1)$, diagram $(j_1)$ is finite but contains logarithms of the scale-ratio $\chi = \tau^2/8\tauf$.
But unlike $(i_1)$, its form is complicated enough that we cannot evaluate it in closed form and must make a $\chi \gg 1$ expansion from the beginning.
Because two-scale problems are difficult, we separate off the part which behaves nontrivially in the UV from the part which behaves nontrivially in the IR, by writing
\begin{equation}
    \label{howtosplit}
    e^{-\tauf(p^2 + k^2 + (p+k)^2)}
    = \Big( e^{-\tauf (p^2 + k^2 + (p+k)^2)} - 1 \Big)
    + 1 \,.
\end{equation}
This rearrangement creates two terms to evaluate, but each involves only a single scale.
They contain canceling logarithmic UV divergences, so they must each be evaluated in the $\MSBAR$ scheme, which is why we have not specialized to $D=4$ yet.
The first term is small unless at least one momentum is of order $\tauf^{-1}$.
that is, it receives its contributions only from UV momentum regions.
We will call this the UV part.
The second term corresponds to the $\MSBAR$-normalized, unflowed $(j)$ diagram.
The nontrivial physics all arises at the $1/\tau$ momentum scale.
We will evaluate it, but it does not contribute to the matching between the flow scheme and the $\MSBAR$ scheme, so it does not have any bearing on \Eq{mainresult}.

First we evaluate the UV part.
If a momentum is large then the vertex $s$ must either be close to 0 or close to $\tau$; the two contribute equally and we will calculate the $s \simeq 0$ case and apply a factor of 2 to cover the other.
This is the region where $k \sim 1/\sqrt{\tauf}$ and $p \sim 1/\tau$.
In this region we can approximate
\begin{equation}
\label{UVapprox}
    2\int_{-\infty}^{\tau/2} ds \: e^{ik_0 s}
    \left( \Theta(-s) - \Theta(s) \right)
    \simeq  \: 2 \int_{-\infty}^\infty dx \: e^{ik_0 s}
    \left( \Theta(-s) - \Theta(s) \right)
    = -\frac{4i}{\kpp}.
\end{equation}
Here $1/\kpp$ means that we take the principal part prescription, eg, $2/\kpp = 1/(k_0+i\epsilon) + 1/(k_0-i\epsilon)$.

The first line of \Eq{j1int} is now of form
\begin{equation}
    \label{j1UVform}
    \int_p \frac{e^{ip_0 \tau}}{p^2} \int_k
    \frac{e^{-\tauf(k^2 + 2k\cdot p)}-1}{k^2 (k+p)^2 \kpp}
\end{equation}
which is small for small $k$ due to cancellations in the numerator, and is small for large $k$ without further positive powers of $k$.
The second line of \Eq{j1int} contains terms with zero, one, or two powers of $k$.
Those terms without powers of $k$, such as $p_0^3$, are small.
Those with one power of $k$ can contribute, but they cancel on angular integration if it is a $k_i$ rather than a $k_0$.
Those with two powers of $k$ cancel on angular integration if we expand \Eq{j1UVform} to lowest order in small $p$, but they contribute if we expand to the next order,
\begin{equation}
\label{largekexpand}
    \frac{e^{-\tauf(k^2 + 2k\cdot p)}-1}{k^2 (k+p)^2 \kpp}
    \simeq \frac{e^{-2\tauf k^2}-1}{k^4 \kpp}
    -2 k\cdot p \frac{\tauf e^{-2\tauf k^2}}{k^4 \kpp}
    -2 k\cdot p \frac{e^{-2\tauf k^2}-1}{k^6 \kpp}.
\end{equation}
Integrating the first term times the $k_0$-terms on the second line of \Eq{j1int} and the second term times the quadratic-in-$k$ terms, one finds:
\begin{equation}
\label{j1UVresult}
    \delta_{(j_1)}^{\mathrm{UV}} G(\tau,\tauf) =
    \frac{g^4 \CA \CR}{16\pi^4 \tau^4}
    \left( \frac{1}{\epsilon} + \ln (2\pi \tauf \tau^2 \mu^4)
    +4 \gamma_E - \frac 53 \right) \times \left\{
    \begin{array}{ll} -3 & E \\ -1 & B \\ \end{array} \right\}.
\end{equation}

Next we address the IR part, or equivalently the $\MSBAR$ renormalized form with the exponential suppression removed.
This time we will have to work with general $s$ values.
We start by rewriting the theta functions to put them in a more convenient form:
\begin{equation}
\label{changethetas}
    \left( \Theta(-s) - \Theta(s) \Theta(\tau-s) + \Theta(s-\tau) \right) = 1 - ( \Theta(s) - \Theta(-s) ) - (\Theta(\tau-s) - \Theta(s-\tau) ).
\end{equation}
The contribution from $\Theta(\tau-s)-\Theta(s-\tau)$
will be the same as the contribution from $\Theta(s)-\Theta(-s)$.
One can see this by reflecting the diagram about $\tau/2$,
or by making the exchange $k\to -k$, $p \to -k-p$.
Therefore we will rewrite the $s$ part of \Eq{j1int} to:
\begin{align}
\label{k0simplify}
    \int_{-\infty}^\infty ds \: e^{ik_0 s}
    \left( \Theta(-s) - \Theta(s) \Theta(\tau-s) + \Theta(s-\tau) \right)
    & \Rightarrow \int_{-\infty}^\infty ds e^{ik_0 s}
    -2 \int_{-\infty}^\infty ds e^{ik_0 s}
    \left( \Theta(s) - \Theta(-s) \right)
    \nonumber \\
    & = 2\pi \delta(k_0) - \frac{4i}{\kpp}.
\end{align}

First we evaluate the contributions arising from $\delta(k_0)$.
These take the form
\begin{equation}
\label{k0deltapart}
    \frac{g^4 \CA \CR}{6} \int_p \frac{e^{ip_0 \tau}}{p^2}
    \int_{\vec k} \frac{1}{\vec k^2 ((\vec p + \vec k)^2+p_0^2)}
    \times \left\{
    \begin{array}{ll}
    -2(D-2) p_0^3  - p_0 p^2 - p_0 \vec k^2 - p_0 (p_0^2 + (\vec p + \vec k)^2) & E \\
    2(D-2) p_0^3 - (D-2) p_0 p^2 + (D-2) p_0 \vec k^2 - (D-2) p_0 (p_0^2 + (\vec p + \vec k)^2) & B \\ \end{array} \right\}.
\end{equation}
We have rearranged the terms in brackets to better cancel against denominators.
The $p^2$ term cancels the $1/p^2$ denominator.
One can then shift $\vec p \to \vec p - \vec k$ to turn the $(p+k)$ denominator into a $p$-denominator.
This leaves the $k$-angular integral $\int_{\vec k} \frac{1}{\vec k^2}$ which is zero in $\MSBAR$.
Similarly, the $(p_0^2 + (\vec p + \vec k)^2)$ term cancels the similar denominator and leaves $\int_{\vec k} \frac{1}{k^2} = 0$.
And the $k^2$ term similarly gives 0.
The only nontrivial expression is the one involving $p_0^3$.
To evaluate it, we write
\begin{align}
\label{k0deltavalue}
    & \frac{2}{3} \int_p \frac{p_0^3 e^{ip_0 \tau}}{p^2}
    \int_{\vec k} \frac{1}{k^2 (p_0^2 + (\vec p + \vec k)^2)}
    \nonumber \\ = & \frac 23 (-i \partial_\tau)^3 
    \int_0^\infty da \: db \: dc \int_{p,\vec k}
    e^{ip_0 \tau} e^{-(a p^2 + b k^2 + c (p_0^2 + (\vec p + \vec k)^2)}
    \nonumber \\ = & \frac 23 (-i \partial_\tau)^3 \int_0^\infty
    da \: db \: dc \: e^{-\tau^2/4(a+c)} \int_{p_0,\vec p,\vec k}
    e^{-(a+c)p_0^2 + \left(a+\frac{bc}{b+c}\right) \vec p^2 + (b+c) \vec k^2}
    \nonumber \\
    = & \frac {2}{3(4\pi)^{7/2}} (-i\partial_\tau)^3 \int_0^\infty da\: db\: dc \:
    e^{-\tau^2/4(a+c)} \frac{1}{(a+c)^{1/2} (ab+ac+bc)^{3/2}}
    \\ \nonumber = &
    \frac{4}{3(4\pi)^{7/2}} (-i \partial_\tau)^3
    \int_0^\infty da \: dc \:
    \frac{e^{-\tau^2/4(a+c)}}{(a+c)\sqrt{ac}}
    \\ \nonumber = &
     \frac{4}{3(4\pi)^{7/2}} (-i \partial_\tau)^3
    \int_0^\infty dy \int_0^1 dx \frac{e^{-\tau^2/4y}}
    {y^{3/2} \sqrt{x(1-x)}}
    \\ \nonumber = & 
     \frac{4}{3(4\pi)^{7/2}} (-i \partial_\tau)^3
    \frac{2 \pi^{3/2}}{\tau}
    = \frac{-i}{8\pi^2} .
\end{align}
In the second line we have replaced $p^0$ factors with $-i\partial_\tau$ factors and exponentiated the denominators.
In the third line we completed all squares, in the fourth line we carried out the Gaussian momentum integrals, in the fifth line we performed the $b$ integral, in the sixth line we defined $y=a+c$ and $x=a/(a+c)$, and the remaining steps are straightforward.

Note importantly that the contribution of the $\delta(k_0)$ term is \textsl{not} zero.
In this respect we appear to disagree with Ref.~\cite{Burnier:2010rp}.

There remains the integral involving $1/k_0$:
\begin{equation}
\label{j1inversek0}
    \frac{2 g^4 \CR \CA}{3} \!\! \int_{pk} 
    \frac{e^{ip_0 \tau}}{\kpp \, p^2 k^2 (p{+}k)^2}
    \left\{ \begin{array}{ll} 
    (2D{-}4) p_0^3 {+} p_0 (p^2{+}k^2{+}(p{+}k)^2)
    {+}3(D{-}2) k_0 p_0^2 {+} \frac{3}{2} k_0 p^2
    {+}(D{-}2) k_0^2 p_0 {+} \frac{k_0}{2} ( k^2 {+} (p{+}k)^2 ) & E \\
    (D{-}2) \left(
    2 p_0^3 {-}  p_0 (p^2 {-} k^2 {+} (p{+}k)^2)
    {+} 3 k_0 p_0^2 {-} \frac{1}{2} k_0 p^2
    {+}  k_0^2 p_0 {+} \frac{1}{2} k_0 (k^2 {-} (p{+}k)^2) \right) & B \\
    \end{array} \right\}
\end{equation}
The terms involving $k_0 p_0^2$, $k_0^2 p_0$, and $p_0 k^2$ must be handled with care as they give rise to $1/\epsilon$ terms.
The sum of the $1/\epsilon$ terms cancel the $1/\epsilon$ term in the UV region, as they must.
Otherwise this part of the calculation is straightforward.

But the most complicated term, and the only one which gives rise to a $\pi^2$ contribution
(besides the $p_0^3 \delta(k_0)$ term we handle above),
is the $p_0^3$ term.
Therefore, and since it is the only place where our results differ from the past literature, we will present it in some detail.
Consider the integral
\begin{equation}
\label{TODO}
    \int_{pk} \frac{ p_0^3 e^{ip_0 \tau}}{\kpp p^2 k^2 (p+k)^2}.
\end{equation}
The first step is to use a trick due to Laine.
The only dependence on the \textsl{sign} of $k_0$ is
in the denominator in the $k_0 (p+k)^2$ term.
We explicitly symmetrize this over relabeling the integration variable $k_0 \leftrightarrow -k_0$:
\begin{align}
\label{TRICK}
    \frac{1}{\kpp ( (p_0+k_0)^2 + (\vec p+\vec k)^2)}
    & = \frac{1}{2} \left(
    \frac{1}{\kpp ( (p_0+k_0)^2 + (\vec p+\vec k)^2)}
    + \frac{1}{(-\kpp) ( (p_0-k_0)^2 + (\vec p+\vec k)^2)}
    \right)
    \nonumber \\ & =
    \left( \frac{ (p_0-k_0)^2 + (\vec p + \vec k)^2
    - (p_0+k_0)^2 - (\vec p - \vec k)^2}
    { \kpp \left( (p_0+k_0)^2 + (\vec p + \vec k)^2 \right) \left( (p_0-k_0)^2 + (\vec p + \vec k)^2 \right)}
    \right)
    \nonumber \\
    & = \frac{-2p_0}
    {\left( (p_0-k_0)^2 + (\vec p+\vec k)^2 \right)
    \left( (p_0+k_0)^2 + (\vec p + \vec k)^2 \right)}.
\end{align}
This cancels the $1/\kpp$ out of the denominator.
Note that this averaging depends crucially on the principal part prescription; otherwise it would not be correct to treat the $k_0$ factors as the same in combining the two denominators.

Using this trick, \Eq{TODO} becomes
\begin{align}
\label{mess1}
    & -2 \int_{pk}
    \frac{p_0^4 e^{ip_0 \tau}}
    {p^2 k^2 (p+k)^2 ((p_0-k_0)^2+(\vec p+\vec k)^2)}
\nonumber \\ = &
-2 \partial_\tau^4 \int_0^\infty da \: db \: dc \: de \:
\int_{pk} e^{ip_0 \tau - p_0^2 (a+c+e) - \vec p^2(a+c+e) - k_0^2(b+c+e)
-\vec k^2(b+c+e) -2p_0k_0(c-e) - 2\vec p\cdot \vec k(c+e)}
\nonumber \\ = &
-2 \partial_\tau^4 \int_0^\infty
\int_{pk} \exp \left( ip_0 \tau - (b+c+e) k^2 -
\frac{ab+(a+b)(c+e)}{b+c+e} \vec p^2
- \frac{ab + (a+b)(c+e)+4ce}{b+c+e} p_0^2 \right)
\nonumber \\ = &
\frac{-2 \partial_\tau^4}{(4\pi)^{7/2}}
\int_{p_0} e^{ip_0 \tau} 
\int_0^\infty \frac{da\: db\: dc\: de}
{ (b+c+e)^{1/2} (ab + (a+b)(c+e))^{3/2}}
\exp \left( \frac{ab+(a+b)(c+e)+4ce}{b+c+e} p_0^2 \right).
\end{align}
Making the change of variables
\begin{align}
    u & = a+b+c+e & a & = xzu \\
    x & = a/(a+b) & b & = (1-x) z u \\
    y & = c/(c+e) & c & = y(1-z) u \\
    z & = (a+b)/(a+b+c+e) & e & = (1-y)(1-z) u \\
    \int_0^\infty da\: db\: cd \: de & = 
    \int_0^1 dx\: dy\: dz \int_0^\infty z(1-z) u^3 du
\end{align}
one finds
\begin{align}
 -\frac{2 \partial_\tau^4}{(4\pi)^{7/2}} \int \frac{dp_0}{2\pi}
    e^{ip_0 \tau}
    \int \frac{dxdydz\; z(1-z) u^3 du}{
      \sqrt{1-xz}(x(1{-}x)z^2+z(1{-}z))^{3/2} u^{7/2}}
    \exp \left( - p_0^2 u \frac{x(1{-}x)z^2 + z(1{-}z) + 4y(1{-}y)
      (1-z)^2}{1-xz} \right)
\end{align}
which suggests a further change of variables:
\begin{align}
  g & \equiv  u \frac{x(1{-}x)z^2 + z(1{-}z) + 4y(1{-}y)
    (1-z)^2}{1-xz}
&
  \frac{u^3 du}{u^{7/2}} & = \frac{\sqrt{1-xz}}{
    \sqrt{x(1{-}x)z^2 + z(1{-}z) + 4y(1{-}y)(1-z)^2}}
  \frac{dg}{\sqrt{g}} \,.
\end{align}
Inserting this change of variables separates the angular integrals from the integral over the overall magnitude:
\begin{align}
 - \frac{2 \partial_\tau^4}{(4\pi)^{7/2}} \int_0^\infty \frac{dg}{g^{1/2}}
  \int \frac{dp_0}{2\pi} e^{ip_0 \tau} e^{-g p_0^2}
  \int_0^1 dxdydz \frac{z(1-z)}{
    (x(1{-}x)z^2+z(1{-}z))^{3/2}
    \sqrt{x(1{-}x)z^2+z(1{-}z) + 4y(1{-}y)(1{-}z)^2}}.
\end{align}
The $(xyz)$-integral is simply $\pi^2/3$.
The $p^0$ and $g$ integrals are straightforward.

Next consider the $(j_2)$ diagram.
Similarly to the $(i_2)$ diagram, there is an integral over the flow time of a flow-dependent vertex, and one fewer power of $k^2$ in the denominator.
Therefore a loop momentum must be large $\sim 1/\sqrt{\tauf}$ for the diagram to contribute.
There are two versions of the diagram, depending on which field strength the flow vertex connects to; they give equal contributions so we will only write one of them and include the other through a factor of 2.
In this case, $k$ must be large and $p$ must be small, and the evaluation is similar to the UV part of diagram $(j_1)$.
Therefore we can make the same treatment of the Wilson-line vertex time as we make in \Eq{UVapprox}.
Doing so, an explicit expression for the diagram becomes:
\begin{equation}
\delta_{(j_2)} G(\tau,\tauf) =
    -\frac{4g^4 \CR \CA}{3} \int_{pk} \int_0^\tauf  ds \frac{e^{ip_0 \tau} e^{-(2\tauf-s)k^2 + s (k+p)^2}}{\kpp p^2 k^2}
\left\{
  \begin{array}{ll}
    6p_0^2(p_0+k_0) + p_0(\vec p^2 + \vec k^2 + (\vec p + \vec k)^2)
    -2k_0 \vec p\cdot \vec k
    & EE \\
    -4 p_0 \vec p \cdot (\vec p + \vec k) & BB \\
  \end{array}
  \right\}.
\end{equation}
We can expand the exponent, $e^{-s(k+p)} \simeq e^{-sk^2}(1-2sk\cdot p)$,
perform the $s$-integration, and drop terms in the brackets with no powers of $k$.
The resulting integrations are straightforward and similar to those for $(i_2)$; the results are presented in Table \ref{TableFlowed}.

Diagram $(j_3)$ is similar, but the vertex occurs on the line which connects to the Wilson line rather than a line attaching to a field strength.
We can again simplify using \Eq{UVapprox}, giving
\begin{align}
    \delta_{(j_3)} G(\tau,\tauf) = -
    \frac{2g^4 \CR\CA}{3} \int_{pk} \int_0^\tauf  ds &
    \frac{e^{ip_0\tau} e^{-sk^2-(2\tauf-s)(p^2+(k+p)^2)}}{\kpp p^2 (p+k)^2}
    \times
    \nonumber \\ & \left\{ \begin{array}{ll}
    6p_0^3+9p_0^2k_0 + 3p_0k_0^2 + p_0(\vec p^2+\vec k^2 + (\vec p+\vec k)^2)
    + k_0(\vec p^2 - \vec p \cdot \vec k) & E \\
    -2(2p_0+k_0)(\vec p \cdot (\vec p + \vec k) & B \\
    \end{array} \right\}.
\end{align}
This is slightly more difficult to compute because we must expand the denominator in $k\gg p$, not just the numerator, but the same techniques apply.
Again, the results have already been presented in Table \ref{TableFlowed}.

Finally we must consider the $(k)$ diagrams, which can all be viewed as self-energy corrections to diagram $(a)$.
The first four represent the vacuum corrections which are also present without flow:  gauge-boson 3-point self-energy, gauge-boson 4-point self-energy, ghost loop, and fermionic loop.
Defining the self-energy as
$\Pi_{\mu\nu}(p) = (g_{\mu\nu} - p_\mu p_\nu/p^2) \Pi(p)$ as usual, the four diagrams contribute
\begin{equation}
    \frac{g^2 \CR}{3} \int_p \frac{e^{ip_0 \tau} e^{-2\tauf p^2}}{p^2} \: \frac{\Pi(p)}{p^2} \left\{
    \begin{array}{ll} -(D-1)p_0^2-\vec p^2 & E \\
    (D-2) \vec p^2 & B \\ \end{array} \right\} 
\end{equation}
which differs from diagram $(a)$ only by the factor of $\Pi(p)/p^2$.
Again, the $p_0$ integral can be pushed into the imaginary plane until it encounters a singularity in the integrand, which first occurs for $p_0 = |\vec p|$.
Therefore $|\vec p| \sim 1/\tau$ to avoid an exponential suppression, which means we can drop the $e^{-2\tauf p^2}$ term and the expression is identical to the unflowed one (up to corrections suppressed by $\exp(-\tau^2/8\tauf)$).
Using
\begin{align}
    \Pi_{123} & = -\frac{g^2 \CA 
    \Gamma \left( 2 - \frac{D}{2} \right)
    \Gamma \left( \frac{D}{2} - 1 \right)
    \Gamma \left( \frac{D}{2} \right)
    (3D-2) \mu^{4-D} (p^2)^{D/2 - 1}}
    { (4\pi)^{D/2} \Gamma(D)},
    &
    \Pi_4 & = \frac{8 g^2 N_f \TR
    \Gamma \left( 2 - \frac{D}{2} \right)
    \Gamma^2 \left( \frac{D}{2} \right)
    \mu^{4-D} (p^2)^{D/2-1}}{(4\pi)^{D/2} \Gamma(D)}
\end{align}
One then generalizes $e^{ip_0 \tau} \to e^{ip\cdot r}$,
expresses the numerator powers of $p_0$ and $p_i$ as
$-i \partial/\partial r_0$ and $-i\partial/\partial r_i$,
performs the $p$-integration using
\begin{equation}
    \int_p e^{ip\cdot r} (p^2)^{D/2-3} =
    \frac{(r^2/4)^{3-D} \Gamma(D-3)}
    {(4\pi)^{D/2} \Gamma(3-D/2)},
\end{equation}
carries out the $r$-derivatives, sets $r_\mu = (\tau,\vec 0)$,
and takes the $D \to 4$ limit.
The result is listed in Table \ref{TableFlowed}.

Next consider $(k_{5,6,7,8})$, where one or more vertices arise from flow.
Each diagram has at least one integral over the flow depth where the vertex arises, contributing a factor of $\tauf$.
Since the momentum $p$ must still be soft $p \sim 1/\tau$, the loop momentum $k$ must be at least $k \sim 1/\sqrt{\tauf}$ to give rise to a leading-order contribution.
This separation of scales means that the problem factorizes into a $k$-integration and a $p$-integration which is identical to diagram $(a)$.
For instance, consider the most challenging diagram, $(k_7)$:
\begin{align}
\label{k7int}
    \delta_{(k_7)} G(\tau,\tauf) & = \frac{g^4 \CR \CA}{3} 
    \int_p \frac{e^{ip_0 \tau} e^{-2\tauf p^2}}{p^2}
    \left\{
    \begin{array}{ll}
         (D-1)p_0^2 + \vec p^2 & E \\
         (D-2) \vec p^2 & B \\
    \end{array}\right\} \times I(p,\tauf)
    \nonumber \\
    I(p,\tauf) & =
    \int_k \int_0^\tauf  ds \:
    \frac{e^{-s(k^2+(p+k)^2-p^2)}}{k^2(p+k)^2}
    \frac{2}{D-1} \left[(6-4D)(p+k)^2 - (D-1) p^2
    +2 (D-2) \frac{(p\cdot k)^2}{p^2} \right].
\end{align}
Expanding in $k \gg p$ and applying the angular averaging
$(p\cdot k)^2 \to p^2 k^2/D$, we find
\begin{align}
    I(p,\tauf) & \simeq
\frac{8 (D-1)}{D} \int_0^\tauf  ds \int_k 
\frac{e^{-2s k^2}}{k^2}
= \frac{8(D-1)}{D} \int_0^\tauf ds
\int_0^\infty da \int_k 
e^{-(2s+a)k^2}
\\ \nonumber & 
= \frac{8(D-1)}{D} \int_0^\tauf ds 
\int_0^\infty da \: \frac{1}{(4\pi)^{D/2} (2s+a)^{D/2}}
= \frac{8(D-1)}{D(4\pi)^{D/2}} \int_0^{\tauf} ds
\int_{2s}^\infty \frac{da}{a^{D/2}}
= \frac{4(D-1)(4\tauf)^{(4-D)/2}}{(4\pi)^{D/2} D \left( \frac{4-D}{2}\right) \left( \frac{D-2}{2}\right)}.
\end{align}
\end{widetext}
The first line of \Eq{k7int} gives \Eq{GEfinal}, and we must take the $D \to 4$ limit of the product of the two.

Note that in the above calculation, we could have directly dropped factors of $p$ in evaluating the vertices and propagators of the self-energy loop, which makes the evaluation still simpler.
Diagrams $(k_5,k_6)$ follow similarly.
Diagram $(k_8)$ vanishes in the $\tau^2 / \tauf \gg 1$ limit.
In evaluating the diagram this occurs because there is no $1/p^2$ arising from any propagator, allowing the $p^0$ contour to be deformed without encountering a singularity.
Physically, the two vertices each arise from flow, and they must be spatially within a distance of $1/\sqrt{\tauf}$ from the two field insertions to avoid an exponential penalty.
Therefore both loop propagators travel a distance $\sim 1/\tau$, and \textsl{both} $p$ and $k$ must be of order $1/\tau$.
Due to the integrations over the flow-depths of the two vertices, the diagram has two powers of $\tauf$ which are not compensated, and the result is suppressed by $\tauf^2/\tau^4$ relative to other diagrams.

Note that the $k_{5\ldots 8}$ diagrams generate the part of the $11 \CA/3$ in the beta function which does not arise from $k_{1\ldots 4}$.
In unflowed perturbation theory, this usually arises from vertex effects.
The $k_{5\ldots 8}$ contributions, arising purely from flow, renormalize the coupling at the scale $\mubar = 1/\sqrt{\tauf}$.
The running between the $\tauf$ and $\tau$ scales instead takes place due to the $(b,i,j)$ diagrams.
The UV divergences in the flow scheme occur because of the $k$ diagrams, while in $\MSBAR$ they occur in $(b,i,j)$.
